\documentclass[prd,nofootinbib,reprint,twocolumn,showpacs,superscriptaddress,floatfix]{revtex4-1}
\pdfoutput=1 
\usepackage{bm, color} 
\usepackage{amssymb,amsfonts,slashed,amsthm,amsmath, graphicx, soul, empheq, cancel}
\usepackage[colorlinks=true, linkcolor=blue, citecolor=blue, urlcolor=blue]{hyperref}
 \usepackage{subcaption}
 \usepackage{cleveref}
 \usepackage{caption}
 
\begin{document}

\newcommand{\vev}[1]{ \left\langle {#1} \right\rangle }
\newcommand{\bra}[1]{ \langle {#1} | }
\newcommand{\ket}[1]{ | {#1} \rangle }
\newcommand{\eV}{ \ {\rm eV} }
\newcommand{\KeV}{ \ {\rm keV} }
\newcommand{\MeV}{\  {\rm MeV} }
\newcommand{\GeV}{\  {\rm GeV} }
\newcommand{\TeV}{\  {\rm TeV} }
\newcommand{\1}{\mbox{1}\hspace{-0.25em}\mbox{l}}
\newcommand{\Red}[1]{{\color{red} {#1}}}

\newcommand{\lmk}{\left(}  
\newcommand{\rmk}{\right)}
\newcommand{\lkk}{\left[}  
\newcommand{\rkk}{\right]}
\newcommand{\lhk}{\left \{ }  
\newcommand{\rhk}{\right \} }
\newcommand{\del}{\partial}  
\newcommand{\la}{\left\langle} 
\newcommand{\ra}{\right\rangle}
\newcommand{\half}{\frac{1}{2}}

\newcommand{\bea}{\begin{array}}
\newcommand{\eea}{\end{array}}
\newcommand{\beq}{\begin{eqnarray}}
\newcommand{\eeq}{\end{eqnarray}}
\newcommand{\eq}[1]{Eq.~(\ref{#1})}

\newcommand{\dd}{\mathrm{d}}
\newcommand{\Mpl}{M_{\rm Pl}}
\newcommand{\mg}{m_{3/2}}
\newcommand{\abs}[1]{\left\vert {#1} \right\vert}
\newcommand{\mphi}{m_{\phi}}
\newcommand{\Hz}{\ {\rm Hz}}
\newcommand{\for}{\quad \text{for }}
\newcommand{\Min}{\text{Min}}
\newcommand{\Max}{\text{Max}}
\newcommand{\Kahler}{K\"{a}hler }
\newcommand{\cphi}{\varphi}
\newcommand{\Tr}{\text{Tr}}
\newcommand{\diag}{{\rm diag}}

\newcommand{\SUf}{SU(3)_{\rm f}}
\newcommand{\Upq}{U(1)_{\rm PQ}}
\newcommand{\Zpq}{Z^{\rm PQ}_3}
\newcommand{\Cpq}{C_{\rm PQ}}
\newcommand{\ubar}{u^c}
\newcommand{\dbar}{d^c}
\newcommand{\ebar}{e^c}
\newcommand{\nubar}{\nu^c}
\newcommand{\Ndw}{N_{\rm DW}}
\newcommand{\Fpq}{F_{\rm PQ}}
\newcommand{\fpq}{v_{\rm PQ}}
\newcommand{\Br}{{\rm Br}}
\newcommand{\Lag}{\mathcal{L}}
\newcommand{\Lqcd}{\Lambda_{\rm QCD}}

\newcommand{\ji}{j_{\rm inf}} 
\newcommand{\jb}{j_{B-L}} 
\newcommand{\M}{M} 
\newcommand{\im}{{\rm Im} }
\newcommand{\re}{{\rm Re} }

\def\lrf#1#2{ \left(\frac{#1}{#2}\right)}
\def\lrfp#1#2#3{ \left(\frac{#1}{#2} \right)^{#3}}
\def\lrp#1#2{\left( #1 \right)^{#2}}
\def\REF#1{Ref.~\cite{#1}}
\def\SEC#1{Sec.~\ref{#1}}
\def\FIG#1{Fig.~\ref{#1}}
\def\EQ#1{Eq.~(\ref{#1})}
\def\EQS#1{Eqs.~(\ref{#1})}
\def\TEV#1{10^{#1}{\rm\,TeV}}
\def\GEV#1{10^{#1}{\rm\,GeV}}
\def\MEV#1{10^{#1}{\rm\,MeV}}
\def\KEV#1{10^{#1}{\rm\,keV}}
\def\blue#1{\textcolor{blue}{#1}}
\def\red#1{\textcolor{blue}{#1}}

\newcommand{\eff}{\Delta N_{\rm eff}}
\newcommand{\neff}{\Delta N_{\rm eff}}
\newcommand{\cc}{\Omega_\Lambda}
\newcommand{\Mpc}{\ {\rm Mpc}}
\newcommand{\Msolar}{M_\odot}

\def\sn#1{\textcolor{red}{#1}}
\def\SN#1{\textcolor{red}{[{\bf SN:} #1]}}
\def\my#1{\textcolor{blue}{#1}}
\def\MY#1{\textcolor{blue}{[{\bf MY:} #1]}}
\def\yn#1{\textcolor{magenta}{#1}}
\def\YN#1{\textcolor{magenta}{[{\bf YN:} #1]}}

\begin{flushright}
\end{flushright}

\title{
How Viable Is a QCD Axion near 10 MeV?
}

\author{Sudhakantha Girmohanta, 
Shota Nakagawa,
Yuichiro Nakai
and Junxuan Xu
\\*[10pt]
{\it \normalsize Tsung-Dao Lee Institute, Shanghai Jiao Tong University, \\
No.~1 Lisuo Road, Pudong New Area, Shanghai, 201210, China} \\*[3pt]
{\it \normalsize School of Physics and Astronomy, Shanghai Jiao Tong University, \\
800 Dongchuan Road, Shanghai, 200240, China}\\
}

\begin{abstract}
There has been an attempt to revive the visible QCD axion at the 10 MeV scale
assuming that it exclusively couples to the first-generation quarks and the electron.
This variant of the QCD axion is claimed to remain phenomenologically viable,
partly due to a clever model construction that induces tree-level pion-phobia and exploits uncertainties inherent
in the chiral perturbation theory.
We confront this model with the cosmological domain wall problem, the quality issue
and constraints arising from the electron electric dipole moment.
It is also pointed out that the gluon loop-generated axion-top coupling can provide a very large contribution
to rare $B$-meson decays, such that the present LHCb data for $B^0 \to K^{*0} e^+ e^-$
rule out the model for the axion mass larger than 30 MeV.
There is a strong motivation for pushing the experimental analysis of $B \to K^{(*)} e^+ e^-$ to a lower $e^+ e^-$ invariant mass window, which will conclusively determine the fate of the model, as its contribution to this branching ratio significantly exceeds the Standard Model prediction.
\end{abstract}

\maketitle
\flushbottom

\section{Introduction
\label{introduction}}
The Standard Model (SM) has achieved incredible success in accounting for the experimental data observed to date,
yet it leaves several theoretical issues unresolved.
The strong CP problem, namely why the CP-violating parameter originating from the topological vacuum nature in QCD
($\bar \theta$) is smaller than $\sim 10^{-10}$, remains one of the outstanding puzzles.\footnote{
It was noted however that, setting $\bar \theta=0$ at some UV scale, the Cabibbo–Kobayashi–Maskawa (CKM) phase
only generates $\bar \theta \sim 10^{-16}$
\cite{Ellis:1978hq}.}
As the viability of the massless up quark has been ruled out from the lattice simulations
\cite{ParticleDataGroup:2022pth, Alexandrou:2020bkd},
the most popular dynamical solution is due to Peccei and Quinn (PQ)
where a global $U(1)_{\rm PQ}$ symmetry is spontaneously broken,
giving rise to a (pseudo-)Nambu-Goldstone boson $a$, the axion~\cite{Peccei:1977hh, Weinberg:1977ma, Wilczek:1977pj}.
Once the axion settles down into the minimum of the potential induced by the strong interaction,
the strong CP problem is dynamically solved. 

The original visible QCD axion, with a decay constant $f_a$ at the electroweak scale,
was soon ruled out as a result of constraints from rare meson decays, and subsequently,
attention was shifted to the invisible axion with a much larger $f_a$ (and smaller mass)~\cite{Kim:1979if, Dine:1981rt}. 
Nevertheless, it has been claimed in Ref.~\cite{Alves:2017avw} that
the exclusion for the axion at the 10 MeV scale is not conclusive,
and a particular pion-phobic variant, where the axion only couples to the first-generation quarks and the electron
by model construction, can still be viable.
This possibility is also motivated by the still indecisive claim of an $e^+ e^-$ excess
in the $^8$Be nuclear de-excitation spectrum~\cite{Krasznahorkay:2015iga}.
Progress has been made toward achieving a realistic UV complete model of this QCD axion variant in Ref.~\cite{Liu:2021wap},
where the PQ symmetry breaking is intertwined with the electroweak symmetry breaking
such that the observed CKM mixing matrix can be reproduced. It is valuable to consider the viability of a QCD axion in this mass range regardless of the indecisive experimental hints and confront it with current experimental and cosmological constraints and discuss possible future probes.

In the present paper, we discuss several new theoretical and experimental consequences of the aforementioned visible QCD axion.
As the decay constant $f_a$ is around GeV for the axion at the 10 MeV scale,
the PQ symmetry breaking is likely to take place in the Universe after inflation.
Further, since this variant necessarily couples to the first-generation quarks, the domain wall number is larger than 1.
Hence, the PQ phase transition generates a string-domain wall network, which results in a catastrophic deviation
from the standard cosmology unless it decays rapidly.
The introduction of a sufficiently large bias term in the form of operators that explicitly break the PQ symmetry 
enables the network to decay.
However, the strength of these operators must be suppressed enough not to spoil the quality of the PQ symmetry
to achieve $\bar \theta \lesssim 10^{-10}$.
We will study the interplay of these two conditions for the axion at the 10 MeV scale. 

The $U(1)_{\rm PQ}$ charge of the up and down quarks must be different to realize the required pion-phobia~\cite{Krauss:1987ud}.
Therefore, in a renormalizable UV completion, one generically needs multiple additional Higgs doublets,
inevitably introducing new CP-violating phases in the scalar potential.
Working with the benchmark model in Ref.~\cite{Liu:2021wap}, we will present those CP-violating phases.
Then, electric dipole moments (EDMs) are generated as a result of the new complex phases. 
We will find that the electron EDM generated at one/two-loop in the model gives the most stringent constraint
on the new physics mass scale. 

Although the tree-level interactions of the visible axion to the higher generation quarks are zero by the model construction,
they are generated via gluon loops.
While these loop generated interactions to the $b$ and $c$-quarks,
constrained by the decays of $J/\Psi$ and $\Upsilon$, have been considered in the literature~\cite{Alves:2017avw,Alves:2020xhf},
the similar coupling to the top quark was neglected, because we cannot produce the corresponding \textit{topped} bound state,
and the top quark decays even before it can hadronize.
Here, however, we will argue that the axion-top coupling generated through a gluon loop gives a dominant new physics contribution
to decays like $B \to K a$ (see Refs.~\cite{Hostert:2020xku, NA62:2023rvm} for constraints from searches of other ultra-rare $K$-decay modes induced by the QCD axion).
In particular, as the present axion variant promptly decays to a $e^+ e^-$ pair to be consistent with the beam dump experiments,
the rare decay mode $B^0 \to K^{*0} e^+ e^-$ can be used to constrain the model.
We show that the contribution of the visible QCD axion to the branching ratio of the decay is much larger than that of the SM.
Then, the current LHCb data~\cite{LHCb:2013pra},
applicable to the invariant mass of the $e^+ e^-$ pair in the range of $30$-$1000$ MeV,
rule out the model for the axion mass $m_a > 30$ MeV.
We emphasize that there is a strong motivation for pushing the analysis for this mode into the lower invariant mass regime,
which will conclusively determine the fate of the present visible axion variant.

The rest of the paper is organized as follows. In Sec.~\ref{sec:review}, we briefly give a review of the QCD axion variant
at the 10 MeV scale and its realistic UV completion.
We discuss the quality and domain wall problems in Sec.~\ref{sec:cosmology}.
Sec.~\ref{sec:eEDM} deals with the induced electron EDM, while
Sec.~\ref{sec:loop} discusses the constraints from the loop-mediated $B$-meson decay.
Finally, we conclude in Sec.~\ref{sec:discussion}.

\section{Review of QCD axion near 10 MeV
\label{sec:review}}

Let us begin with a review of the QCD axion variant with mass of $\mathcal{O}(10)\MeV$ discussed in
Refs.~\cite{Alves:2017avw,Alves:2020xhf,Liu:2021wap}.
It had been widely believed that such a visible axion model can be easily excluded by beam dump experiments
\cite{Donnelly:1978ty} or measurements of rare heavy meson decays
\cite{Wilczek:1977zn,Hall:1981bc}.
However, the authors of Ref.~\cite{Alves:2017avw} claimed
a possible way out of those severe constraints on the model, which is composed of three points:
(i) the axion must couple to only the first-generation of the SM fermions,
(ii) the axion is required to be pion-phobic~\cite{Krauss:1987ud},
and (iii) there exist large uncertainties in theory and experiment.
For point (i), only the axion couplings to the first-generation quarks ($u,d$) are allowed
to avoid exotic decays of the heavy quarkonia, e.g. $J/\psi$ and $\Upsilon$
\cite{ARGUS:1986ytx,Hsueh:1992ex,E760:1992pgt,BES:1995wyo,CLEO:2005cdx}.
The axion must couple to the electron as otherwise it would become too long-lived and would
have been already detected by beam dump experiments (the lifetime is bounded as $\tau\lesssim10^{-13}\sec$) \cite{CHARM:1985anb,Bjorken:1988as,Davier:1989wz,Blumlein:1990ay}.
The axion coupling to the muon is prohibited by the measurement of the muon magnetic dipole moment
unless the decay constant $f_a$ is larger than the electroweak scale.
Regarding point (ii), pion-phobia enables the axion-pion mixing to be suppressed,
leading to the tiny rate of an exotic decay of the pion.
The SINDRUM collaboration~\cite{SINDRUM:1986klz} searches for the process $\pi^+\rightarrow e^+\nu_e a(\rightarrow e^+e^-)$
and provides a constraint on the branching ratio, resulting in the upper bound on the axion-pion mixing. 
To realize pion-phobia, one can take special PQ charge assignments of $u$ and $d$,
\beq
\frac{Q_u}{Q_d}=2,
\label{Eq:rqud}
\eeq
leading to the cancellation due to the mass ratio between $u$ and $d$.  
For point (iii), the axion-$\eta$ meson mixing induces a $K^+ \to \pi^+ a$ decay, however, the inherent uncertainties in the chiral perturbation theory have been used to argue that the QCD axion variant remains viable~\cite{Alves:2017avw}. Note that Eq.~\eqref{Eq:rqud} only realizes tree-level pion-phobia, and the effective PQ charges receives corrections due to the renormalization group evolution (RGE)~\cite{Chala:2020wvs, Bauer:2020jbp}. For example, including the dominant contribution from a top-induced vertex correction, the effective PQ charges for the $u$, $d$ becomes 
\begin{align}
    \nonumber
    Q_u^{(1)} &= Q_u^0 -6 \left( \frac{y_t}{4\pi}\right)^2 \ln\left( \frac{\Lambda_{\rm UV}}{\mu} \right) \ , \\
    Q_d^{(1)} &= Q_d^0 + 6 \left( \frac{y_t}{4\pi}\right)^2 \ln\left( \frac{\Lambda_{\rm UV}}{\mu} \right) \ ,
\end{align}
where the running from $\Lambda_{\rm UV}$ to low-energy scale $\mu$ is considered, and $y_t$ is the Yukawa coupling for top, and $Q_u^0=2$, $Q_d^0=1$ are the tree-level PQ charge assignments. The resulting change in axion-pion mixing angle is less than its current uncertainty stemming from $m_u/m_d$ measurement. Together with the uncertainties in the higher order chiral perturbation theory, pion-phobia can be realized including RGE effects, perhaps in a slightly different parameter space~\cite{DiLuzio:2022tyc}.

For other requirements, we should mention the bound from the electron magnetic moment, $(g-2)_e$.
The axion-electron and -photon couplings induce loop contributions to $(g-2)_e$ and
thus receive a constraint from the current measurements.\footnote{
There is a discrepancy of values of $(g-2)_e$ calculated by using values of the fine-structure constant
measured by two independent experiments.
While the cesium recoil experiment
\cite{Parker:2018vye} leads to a smaller value of $(g-2)_e$,
the experiment using the rubidium atom \cite{Morel:2020dww} leads to a larger value.
The recent measurement of $(g-2)_e$ \cite{Fan:2022eto} does not solve the tension.}
In addition, the axion coupling to the electron is constrained by KLOE \cite{Anastasi:2015qla} and NA64 experiments
\cite{NA64:2019auh,NA64:2020xxh}.
Taking account of both bounds, we obtain the allowed range of the PQ charge of the electron, $1/3\lesssim Q_e\lesssim2$.

\begin{table}[tp]
\vspace{0mm}
\centering
\begin{tabular}{c|c|c|c|c|c|c|c}
& $H$ & $H_u$ & $H_d$ & $H_e$ & $\Phi_u$ & $\Phi_d$ & $\Phi_e$ \\
\hline
$SU(2)_L$ & $\bm 2$ & $\bm 2$ & $\bm 2$ & $\bm 2$ & $\bm 1$ & $\bm 1$ & $\bm 1$ \\
$U(1)_Y$  & $1/2$ & $-1/2$ & $1/2$ & $1/2$ & $0$ & $0$ & $0$ \\
$U(1)_{\rm PQ}$ & $0$ & $-Q_u$ & $-Q_d$ & $-Q_e$ & $-Q_u$ & $-Q_d$ & $-Q_e$ \\ 
\end{tabular}
\vspace{1mm}
\caption{The charge assignments of scalar fields.}
\label{tab:charge}
\end{table}

We now consider a UV model to satisfy the requirements, following the discussion of Ref.~\cite{Liu:2021wap}.
The model introduces four doublet Higgs fields $H,H_u,H_d,H_e$ and three SM-singlet scalar fields $\Phi_u,\Phi_d,\Phi_e$.
Here, $H$ corresponds to the usual SM Higgs doublet.
The charge assignments of these scalar fields are summarized in Tab. \ref{tab:charge},
where we define the PQ charges of the right-handed up quark $u_R$, down quark $d_R$ and electron $e_R$ as $Q_u,Q_d$
and $Q_e$, respectively. 
In the following discussion, we set them as
\beq
Q_u=2, \quad Q_d=1, \quad Q_e=\frac{1}{n} ,
\eeq
with $n=2,3$.
This choice makes the axion pion-phobic and evades the bound on $Q_e$ conservatively.
The color anomaly is estimated as $(Q_u+Q_d)/2$, and the domain wall number is three. 
Yukawa interactions with the new Higgs doublets and the SM one are respectively given by
\beq
\Lag_{\rm PQ}^Y &=& -\sum_{i=1,2,3}\lmk \bar{Q}^{i}Y_u^{i1}H_uu_R^{1}+\bar{Q}^{i}Y_d^{i1}H_dd_R^{1}\right.\nonumber\\
&+& \left.\bar{L}^{i}Y_e^{i1}H_ee_R^{1}\rmk +{\rm h.c.},\\[1ex]
\Lag_{\rm SM}^{Y}&=& -\sum_{i=1,2,3}\sum_{j=2,3}\lmk \bar{Q}^{i}Y_u^{ij}\tilde{H}u_R^{j}+\bar{Q}^{i}Y_d^{ij}Hd_R^{j}\right.\nonumber\\
&+& \left.\bar{L}^{i}Y_e^{ij}He_R^{j}\rmk +{\rm h.c.},
\eeq
where $Q,L$ respectively denote the left-handed quark and lepton, the indices $i,j$ represent the generation 
and $\tilde{H} \equiv i \sigma_2 H^\ast$.
The first Lagrangian involves the first generation right-handed fermions,
while the second one is relevant only to the second and third generation right-handed fermions. 
With these Yukawa interactions, the CKM matrix can be derived successfully
\cite{Liu:2021wap}.

The potential terms of the scalar fields to break the electroweak and PQ symmetries include
\beq
V_{\rm PQ} &=& \bigl( A_{1}H H_u\Phi_u^* +A_2HH_d^\dagger\Phi_d+ A_{3}H_eH^\dagger\Phi_e^* \nonumber\\
&+&  A_4\Phi_u^*\Phi_d^2 +A_5\Phi_d^*\Phi_e^n \bigr) + \rm {h.c.},\label{scalarV1}\\[2ex]
V_{\rm dia} &=& \sum_{\Psi}-\mu_\Psi\Psi^\dagger\Psi +\lambda_\Psi(\Psi^\dagger\Psi)^2 .\label{scalarV2} 
\eeq
Here, $\Psi = H,H_u,H_d,H_e,\Phi_u,\Phi_d,\Phi_e$ and
$A_1,A_2,A_3,A_4$ and $\mu_\Psi$ are the parameters with mass dimension one and two, respectively.
The mass dimension of the parameter $A_5$ depends on the choice of $n$, and $\lambda_\Psi$ is a dimensionless coupling.
Although other PQ-symmetric operators can be written down, they are not relevant for the PQ mechanism and omitted here.
However, in \SEC{sec:eEDM}, we will show that such extra terms induce sizable contributions to EDMs.
In terms of seven neutral pseudoscalars, $\vec{\Phi}^I=(h^I,-h_u^{I},h_d^I,h_e^I,\phi_u^I,\phi_d^I,\phi_e^I)^T$
where we take $-h_u^I$ for $\tilde{H}_u$ whose charge is the same as that of $H_{d,e}$, the physical axion degree of freedom is written as
\beq
a&\simeq& \frac{1}{\sqrt{\sum_fQ_f^2(v_f^2+v_{\Phi_f}^2)} }\biggl( -\sum(-1)^fQ_fv_f^2/v,-Q_uv_u, \nonumber\\
&&Q_dv_d, Q_ev_e, Q_uv_{\Phi_u}, Q_dv_{\Phi_d}, Q_ev_{\Phi_e} \biggr) \, \vec{\Phi}^I \nonumber \\[1ex]
&\equiv& \vec{G}_{\rm PQ}^T\vec{\Phi}^I,
\eeq
where $v \, (\simeq 246\GeV), v_u, v_d, v_e, v_{\Phi_u}, v_{\Phi_d}, v_{\Phi_e}$ denote vacuum expectation values (VEVs)
of $H,H_u,H_d,H_e, \Phi_u,\Phi_d$ and $\Phi_e$, respectively, and
$(-1)^f=+1$ for $f=d,e$ and $(-1)^f=-1$ for $f=u$. 
In estimating the prefactor, we have used $v_f\ll v$.
The axion-fermion coupling is then given by
\beq
\Lag_{aff}&=&\sum_{f=u,d,e}\frac{m_f}{v_a}Q_fia\bar{f}\gamma_5f\nonumber\\
&-& \frac{\sum_f(-1)^fQ_fv_f^2}{v^2}\sum_{F={\rm 2nd,3rd}} \frac{m_F}{v_a}ia\bar{F}\gamma_5F \, ,
\eeq
with the PQ breaking scale, $v_a\equiv \sqrt{\sum_fQ_f^2(v_f^2+v_{\Phi_f}^2)}$. 
Note that we have the axion couplings to the 2nd and 3rd generation fermions, but they are suppressed by $(v_f/v)^2$.~\footnote{The axion coupling to the muon through Higgs mixing effects is tiny. Quantitatively, the axion contribution to muon magnetic moment $(g-2)_\mu \lesssim 10^{-20}$.} In \SEC{sec:loop}, we will point out extra contributions to those couplings induced by gluon loops.

The five physical pseudoscalar bosons (other than the axion and the Nambu-Goldstone mode absorbed by the $Z$ boson)
need to be heavy enough to evade the current experimental searches.
In Ref.~\cite{Liu:2021wap}, the authors take the magnitudes of the VEVs and the $A$-term coefficients as
\beq
&&A_4\gg A_{k (=1,2,3)}\simeq 20\GeV  , \nonumber \\[1ex]
&&v_{f (=u,d,e)} \simeq 20\MeV  , \quad v_{\Phi_{f (=u,d,e)}} \simeq 1\GeV .
\label{parameter}
\eeq
The value of $v_{\Phi_f}$ sets the PQ breaking scale to $\mathcal{O}(1)\GeV$ ($m_a = \mathcal{O}(10)\MeV$). 
Then, the three pseudoscalars are approximately given by linear combinations of $h_f^I$s,
and their masses are estimated as $A_kvv_{\Phi_f}/v_f\sim(300\GeV)^2$.
Looking at the fourth term in \EQ{scalarV1}, one massive mode is described as a linear combination of $\Phi_u$ and $\Phi_d$,
and its mass is given by $A_4v_{\Phi_f}\gg(1\GeV)^2$.
We can take a large value of $A_4$ to make this mode sufficiently heavy.
For $n=2$, $A_5$ is dimensionful, and hence the last pseudoscalar mode can be heavy by taking a large value of $A_5$.
On the other hand, for $n=3$, $A_5$ is a dimensionless constant,
so that the last mode has a relatively small mass of $A_5v_{\Phi_f}^2\sim(1\GeV)^2$. This mode is dominantly composed of $\Phi_e$ without other $A$-term interactions, and then mainly couples to the electron. The coupling strength is estimated as $|Y_e|v_f/v_{\Phi_f}\simeq m_e/v_{\Phi_f}\simeq5\cdot10^{-4}$,
and constrained by dark photon searches, such as BaBar \cite{BaBar:2014zli} for the mass range of $1-10\GeV$.
Hence, the case of $n=3$ is marginally consistent with the current bound.
In addition to those pseudoscalars, there are seven CP-even scalars in the model.
First, we have the SM-like Higgs boson with mass-squared of $\sim \lambda_H v^2$.
Then, as in the case of the pseudoscalar modes,
three modes obtain masses of $\sim A_kvv_{\Phi_f}/v_f$ and two modes have masses of $\sim A_4v_{\Phi_f}$. 
For $n=2$, taking a large value of $A_5$, all the CP-even scalars are sufficiently heavy.
For $n=3$, there remains a light mode whose mass is of $\mathcal{O}(1)\GeV$, and the similar experimental constraint
as that of the pseudoscalar is applied to this mode.\footnote{The reported $e^+e^-$ excess in the de-excitation of $^8$Be and $^4$He  \cite{Krasznahorkay:2015iga,Krasznahorkay:2019lyl} and $(g-2)_e$ can be simultaneously explained by this axion model for $n=2,3$~\cite{Alves:2020xhf,Liu:2021wap}, we mainly consider the case of $n=2,3$ in the present paper for comparison. However, as the uncertainty in fine structure measurements is larger than the uncertainty in $(g-2)_e$, disregarding the $(g-2)_e$ discrepancy explanation, the $n=1$ case can be allowed, when the model can be more minimal, composed of five Higgs fields.}
We assume the magnitudes of Eq.~\eqref{parameter} in the following discussion.

\section{Quality and Cosmology
\label{sec:cosmology}}

The PQ mechanism can solve the strong CP problem dynamically under the assumption that the PQ symmetry is well preserved.
However, the PQ symmetry is introduced as a global symmetry, and we naturally expect Planck-suppressed operators
breaking the PQ symmetry explicitly
\cite{Dine:1986bg, Barr:1992qq, Kamionkowski:1992mf, Kamionkowski:1992ax, Holman:1992us, Kallosh:1995hi, Carpenter:2009zs, Carpenter:2009sw, Girmohanta:2023ghm}.
For invisible axion models with large $f_a$, they destroy the PQ solution, which is known as the axion quality problem.
Here, we discuss whether the similar problem exists or not in the visible axion model with $f_a = \mathcal{O}(1) \, \rm GeV$.
We also explore a possibility that those PQ violating operators provide bias terms to solve the domain wall problem.

\subsection{Explicit PQ breaking operators}

For the visible QCD axion model presented in Sec.~\ref{sec:review}, PQ-breaking higher-dimensional operators
that are the most dangerous for the PQ solution are
\beq
\Lag_{\cancel{\rm PQ}} = \sum_{f=u,d,e} \!\!\!\!\!\!\!\!\!\!\!  &&\lmk g_1\frac{(HH^\dagger)^2\Phi_f}{\Mpl} + g_2\frac{HH^\dagger\Phi_f^3}{\Mpl}\right.\nonumber\\
&+& g_3\left. \frac{\Phi_f^5}{\Mpl} + \cdots \rmk + {\rm h.c.},
\label{LPQV}
\eeq
where each $g_i = |g_i|e^{i\delta_i}$ $(i=1,2,3)$ is a complex coupling constant with a phase shift $\delta_i$ and is assumed to be independent of $f$ for simplicity. 
Here, we have omitted to write down combinations such as $\Phi_u^3\Phi_d^2$, which will be taken into account in the final results. The three terms in \EQ{LPQV} respectively generate the following axion potential terms:
\beq
V^{(1)}_{\cancel{\rm PQ}} &=& -\frac{|g_1|}{2\sqrt{2}}\frac{v^4}{\Mpl} \sum_f v_{\Phi_f}\cos\lmk Q_f\frac{a}{v_a} +\delta_1 \rmk, \label{PQV1}\\[1ex]
V^{(2)}_{\cancel{\rm PQ}} &=& -\frac{|g_2|}{2\sqrt{2}}\frac{v^2}{\Mpl} \sum_f v_{\Phi_f}^3\cos\lmk 3Q_f\frac{a}{v_a} +\delta_2 \rmk,\label{PQV2}\\[1ex]
V^{(3)}_{\cancel{\rm PQ}} &=& -\frac{|g_3|}{2\sqrt{2}} \sum_f \frac{v_{\Phi_f}^5}{\Mpl}\cos\lmk 5Q_f\frac{a}{v_a} +\delta_3 \rmk. \label{PQV3}
\eeq
They shift the axion VEV from the CP conserving minimum, which is set by the potential from QCD nonperturbative effects,
$V_a=m_a^2f_a^2(1-\cos(a/f_a))$. Here the axion mass $m_a$ is given by
\beq
m_a=\frac{\sqrt{m_um_d}}{m_u+m_d}\frac{m_\pi f_\pi}{f_a} \, ,
\label{Eq:QCD_relation}
\eeq
where the decay constant of the axion is defined as $f_a\equiv v_a/N_{\rm DW}$ with domain wall number
$N_{\rm DW}(=Q_u+Q_d=3)$, and $m_\pi\simeq 135\MeV$ and $f_\pi\simeq92\MeV$ denote the pion mass and decay constant, respectively.
For the term $V_{\cancel{\rm PQ}}^{(1)}$, the shift of the axion VEV is evaluated as
\beq
\frac{|\langle a\rangle|}{f_a} \simeq \frac{|g_1|}{2\sqrt{2}N_{\rm DW}}\frac{v^4}{\Mpl m_a^2f_a^2}\sum_fQ_fv_{\Phi_f}\sin\delta_1.
\eeq
Those corresponding to $V_{\cancel{\rm PQ}}^{(2)}$ and $V_{\cancel{\rm PQ}}^{(3)}$ can be estimated similarly.

\begin{figure*}[!t]
\includegraphics[width=8cm]{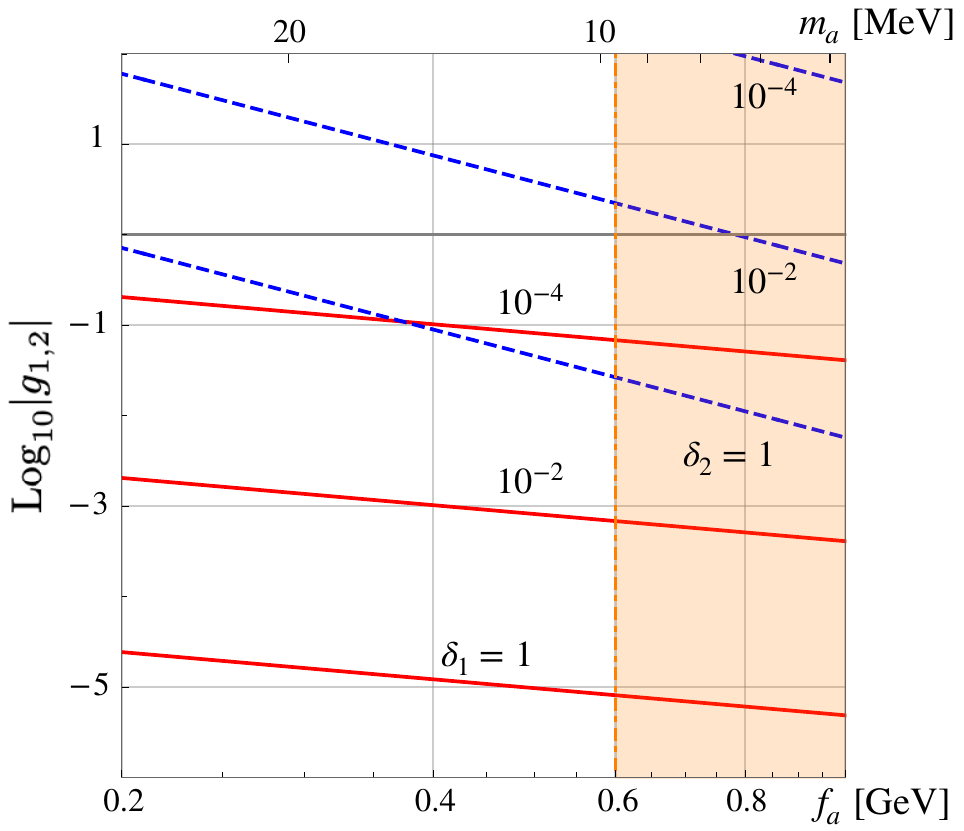}
\includegraphics[width=8cm]{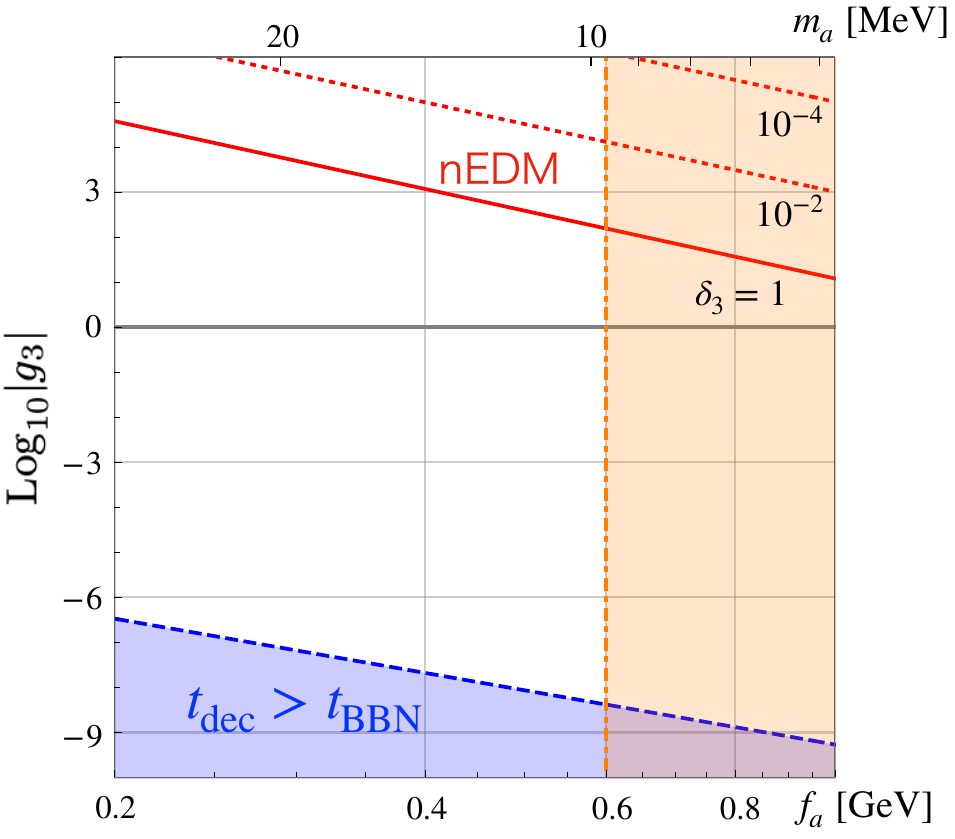}
\caption{
{\it Left panel \hspace{0.1mm}}: upper bounds on the coupling constants $|g_{1,2}|$ from the neutron EDM measurement
as functions of $f_a$ (lower axis) or $m_a$ (upper axis) for the case of $n=2$.
The red solid and blue dashed lines represent the cases of $\delta_{1,2}=1,10^{-2},10^{-4}$ for \EQ{PQV1} and \EQ{PQV2}, respectively.
{\it Right panel \hspace{0.1mm}}: bounds on the coupling constant $|g_3|$ of \EQ{PQV3} from the neutron EDM measurement for $n=2$.
The red solid and dotted lines denote the bounds for $\delta_3=1,10^{-2},10^{-4}$.
In the blue shaded region, the domain wall survives after BBN. 
In both panels, the orange shaded regions are excluded by beam dump experiments
\cite{Riordan:1987aw}.
}
\label{fig:g-fa}
\end{figure*}

The current neutron EDM measurement provides an upper bound on $\bar{\theta}$
\cite{Baker:2006ts,Pendlebury:2015lrz}, such that
$|\langle a \rangle|/f_a \lesssim 10^{-10}$.
This bound can be read out as upper bounds on the coupling coefficients $|g_i|$.
Considering each operator as the dominant one, we obtain
\beq
|g_1| &\lesssim& 
8\times10^{-6}  \lmk\frac{m_a}{10\MeV}\rmk,\\[1ex]
|g_2| &\lesssim& 
0.03 \lmk\frac{\Delta^{(2)}}{10}\rmk^{-1}\lmk\frac{3}{N_{\rm DW}}\rmk^2\lmk\frac{m_a}{10\MeV}\rmk^3,\\[1ex]
|g_3| &\lesssim&  
200 \lmk\frac{\Delta^{(3)}}{100}\rmk^{-1}\lmk\frac{3}{N_{\rm DW}}\rmk^4\lmk\frac{m_a}{10\MeV}\rmk^5. 
\eeq
Here, we use $f_a\simeq\sqrt{\sum Q_f^2 v_{\Phi_f}^2}/N_{\rm DW}$ and take $n=2$ and $\delta_i=1$. 
Since we expect $\mathcal{O}(10)$ and $\mathcal{O}(100)$ number of operators in Eq.~\eqref{LPQV} giving similar contributions to \EQ{PQV2} and \EQ{PQV3}, respectively,
the factors of $\Delta^{(2)}=\mathcal{O}(10)$ and $\Delta^{(3)}=\mathcal{O}(100)$ are included for
the bounds on $|g_2|$ and $|g_3|$.
The results show that fine-tuning of $\mathcal{O}(10^{-5})$ is required for the PQ quality
in the presence of the coupling $g_1$.
The $f_a$ (or $m_a$) dependence of bounds on $|g_i|$ is also summarized in \FIG{fig:g-fa}.
The results for $n=3$ are not significantly different from those of $n=2$.

While, compared to the invisible QCD axion models, the PQ quality for the visible axion at the 10 MeV scale is much better,
the complete PQ quality requires removal of higher-dimensional operators involving the SM Higgs.
This conclusion comes from the fact that the PQ breaking scale is below the electroweak scale.
A small axion decay constant is not sufficient to achieve the high quality of the PQ symmetry.
However, this issue might be addressed by a further model-building.

\subsection{Domain wall problem}

Spontaneous breaking of the PQ symmetry after inflation
generates topological defects which can give a non-negligible impact on the evolution of the Universe. At the PQ phase transition, cosmic strings are formed through the Kibble mechanism \cite{Kibble:1976sj,Kibble:1982dd},
and when the axion acquires a potential from nonperturbative effects of QCD,
domain walls are produced and the string-wall system appears.
Its cosmological consequence depends on the domain wall number $N_{\rm DW}$ of a model.
If $N_{\rm DW}=1$, the string-wall system is unstable and collapses with a short lifetime. 
On the other hand, if $N_{\rm DW} > 1$, it is stable and may quickly dominate the Universe.
This is known as the cosmological domain wall problem (see Refs.~\cite{Beyer:2022ywc,Dine:2023qsq} for recent reviews).
One possible solution to the domain wall problem is to introduce an explicit PQ breaking operator (often called bias term), which lifts up the degenerate vacua and makes the string-wall system unstable
\cite{Vilenkin:1981zs,Gelmini:1988sf,Larsson:1996sp}. 
However, for invisible axion models, the solution is in tension with the quality of the PQ symmetry,
and one has to adjust the axion potential from the explicit breaking operator so that its minimum is aligned with
that of the potential from nonperturbative effects of QCD
\cite{Kawasaki:2014sqa}.

The situation is improved considerably for the visible axion model with $f_a = \mathcal{O}(1) \, \rm GeV$. In order to solve the domain wall problem, it is sufficient to find a condition that the domain wall collapses
through a PQ-breaking operator before the Big Bang Nucleosynthesis (BBN) so as not to destroy the light elements.
The PQ-breaking operator makes an energy difference $\Delta V$ between the vacua,
which induces a volume pressure on the domain wall.
The wall becomes unstable when the volume pressure is comparable to the domain wall tension:
\beq
\Delta V \sim \frac{\beta\sigma_{\rm wall}}{t_{\rm dec}} \, .
\eeq
Here, $\sigma_{\rm wall}\simeq 9.25m_af_a^2$ is the surface density of the domain wall \cite{Huang:1985tt,Hiramatsu:2012sc}, $\beta$ is a constant which is $\mathcal{O}(1)$ by dimensional analysis,
and $t_{\rm dec}$ is the cosmic time when the domain wall decays. The potential difference between the origin and its neighbor vacuum is approximately given by
\beq
\Delta V \sim |V_{\cancel{\rm PQ}}(a=2\pi v_a/N_{\rm DW})-V_{\cancel{\rm PQ}}(0)| \, .
\eeq
For $V_{\cancel{\rm PQ}}^{(3)}$ in Eq.~\eqref{PQV3},
the constraint from the domain wall lifetime is shown in the blue shaded region of the right panel of \FIG{fig:g-fa}. Although we take $n=2$ here, a similar result can be obtained for $n=3$.  
We can see that the constraint is still satisfied without causing the axion quality problem.
For $V_{\cancel{\rm PQ}}^{(1,2)}$d in Eqs.~\eqref{PQV1}, \eqref{PQV2}, the domain wall decays even faster,
and the constraint does not appear in the left panel of \FIG{fig:g-fa} although there is a quality issue for those operators.
In conclusion, if we can somehow suppress higher-dimensional operators involving the SM Higgs
which lead to $V_{\cancel{\rm PQ}}^{(1,2)}$ but allow the third term of Eq.~\eqref{LPQV} leading to Eq.~\eqref{PQV3}
by model-building, the domain wall problem is solved with the high-quality of the PQ symmetry.

\section{Electron EDM
\label{sec:eEDM}}

\begin{figure*}[t]
    \centering
    \begin{subfigure}{0.32\linewidth}
        \includegraphics[width=0.8\textwidth]{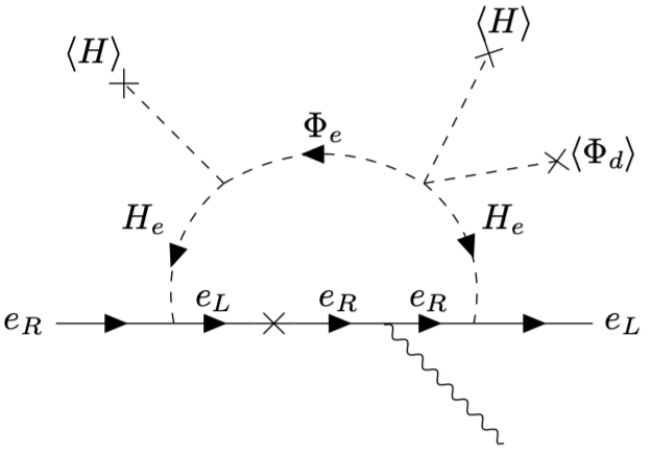}
        \caption{One-loop diagram in the case of $n=2$}
        \label{sub1}
    \end{subfigure}
    \begin{subfigure}{0.32\linewidth}
        \includegraphics[width=0.8\textwidth]{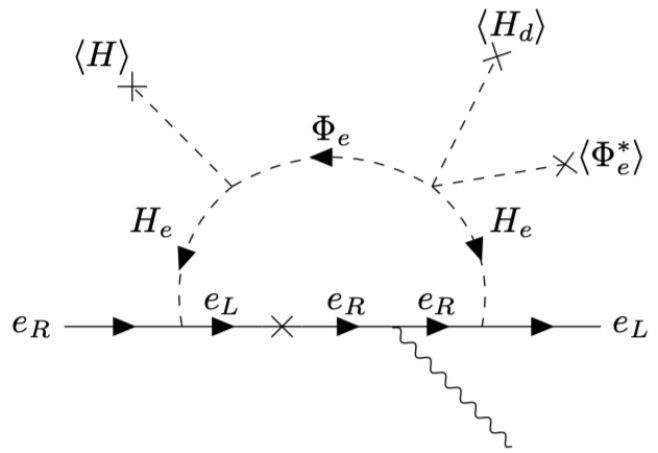}
        \caption{One-loop diagram in the case of $n=3$}
        \label{sub2}
    \end{subfigure}
    \begin{subfigure}{0.32\linewidth}
        \includegraphics[width=0.8\textwidth]{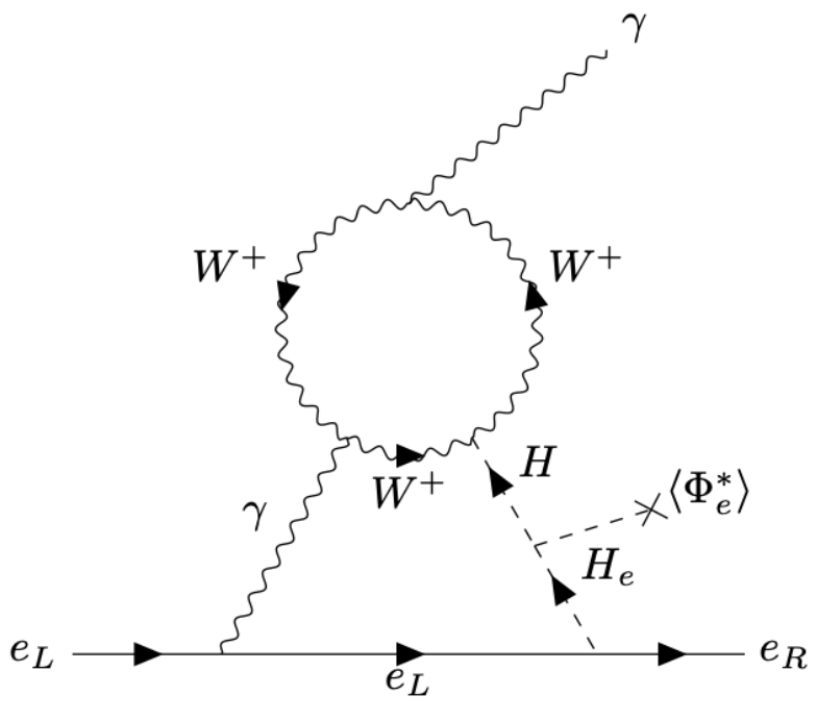}
        \caption{Two-loop Barr-Zee diagram with a $W$-loop}
        \label{sub3}
    \end{subfigure}
    \caption{Feynman diagrams for dominant contributions to the electron EDM. The arrows in the diagram represent charge flow.}
    \label{1-loop eEDM n=2}
\end{figure*}

The model contains three Higgs doublet fields $H_f$ ($f=u,d,e$) and SM-singlet scalars $\Phi_f$ as well as the SM Higgs field.
According to the charge assignment summarized in Tab.~\ref{tab:charge},
many scalar potential terms other than those presented in Eqs.~\eqref{scalarV1}, \eqref{scalarV2} are actually allowed
by the SM gauge and PQ global symmetries.
In the case of $n=2$, all the possible (off-diagonal) scalar potential terms are
\begin{align}
    V^{(n=2)}_{\rm scalar} &= A_1H H_u\Phi_u^*+A_2HH_d^\dagger\Phi_d+ A_{3}H_eH^\dagger\Phi_e^* \notag\\
    &+A_{4}\Phi_u^*\Phi_d^2+A_{5}\Phi_d^*\Phi_e^2 +A_6H_dH_e^\dagger\Phi_e^*\notag\\
    &+B_1HH_u{\Phi_d^*}^2+B_2HH_d^\dagger\Phi_u\Phi_d^*+B_3 HH_d^\dagger\Phi_e^2\notag\\
    &+B_4 H^\dagger H_e\Phi_d^*\Phi_e+B_5 H_uH_e\Phi_u^*\Phi_e^*+B_6H_uH_d\Phi_u^*\Phi_d^*\notag\\
    &+B_7H_d^\dagger H_e\Phi_d\Phi_e^*+B_{8}\Phi_u\Phi_d^*{\Phi_e^*}^2+B_9 H_e^2H^\dagger H_d^\dagger\notag\\
    &+\rm{h.c.}\, ,
\end{align}
where $A_j$ ($j=1,2,\cdots,6$) and $B_k$ ($k=1,2,\cdots,9$) are parameters with mass dimension one and
dimensionless coupling constants, respectively.
All the Higgs doublets and singlets get VEVs.
Those parameters and VEVs are generally complex.
Field redefinition rotates away some of the phases but is not sufficient to remove all, leading to observable phases.

The physical complex phases violate CP symmetry and lead to EDMs.
Fig.~\ref{sub1} describes a one-loop diagram generating the electron EDM.
The diagram includes the interactions, $A_{3}^*H_e^\dagger H\Phi_e$ and $B_{4}^*HH_e^\dagger\Phi_d\Phi_e^*$.
The contribution to the electron EDM is then estimated as
\begin{align}
    \frac{d_e}{e} \sim \frac{\theta}{16\pi^2} | A_{3}^*B_{4}^*v^2v_{\Phi_d} | \frac{m_e}{M^6}\, .\label{n=2EDM}
\end{align}
Here, $\theta$ and $M$ denote the relevant physical phase and a representative mass scale of particles in the loop,
respectively.
The current experimental upper limit on the electron EDM is $d_e\leq4.1\times10^{-30} e \,\rm cm$~\cite{Roussy_2023}.
With our choice of Eq.~\eqref{parameter},
the constraint can be satisfied for the mass scale,
\begin{align}
    M\gtrsim 500 \, \rm GeV\, ,\label{bound1}
\end{align}
where we have assumed the phase $\theta$ and the dimensionless coupling $B_4$ are of order one.
As we have seen in Sec.~\ref{sec:review}, all the (pseudo-)scalar particles inside the loop
can have masses of $\mathcal{O}(100) \, \rm GeV$ in the case of $n=2$.
Therefore, the condition \eqref{bound1} is marginally satisfied.

In the case of $n=3$, the possible (off-diagonal) scalar potential terms allowed by the symmetries are
\begin{align}
    V^{(n=3)}_{\rm scalar} &= A_{1}H H_u\Phi_u^* +A_2HH_d^\dagger\Phi_d+ A_{3}H_eH^\dagger\Phi_e^* \notag\\
    &+A_{4}\Phi_u^*\Phi_d^2+A'_5\Phi_d^*\Phi_e^3+B_1 HH_u{\Phi_d^*}^2\notag\\
    &+B_2HH_d^\dagger\Phi_u\Phi_d^* + B_5 H_uH_e\Phi_u^*\Phi_e^*+B_6 H_uH_d\Phi_u^*\Phi_d^* \notag\\
    &+ B_7 H_d^\dagger H_e\Phi_d\Phi_e^* + B_{10} H_dH_e^\dagger{\Phi_e^*}^2+\rm{h.c.}\, ,
\end{align}
where $A'_5$ and $B_{10}$ denote dimensionless coupling constants.
The diagram shown in Fig.~\ref{sub2} contains the interactions,
$A_{3}^*H_e^\dagger H\Phi_e$ and $B_{10} H_dH_e^\dagger{\Phi_e^*}^2$.
The contribution to the electron EDM is then estimated as
\begin{align}
    \frac{d_e}{e} \sim\frac{\delta}{16\pi^2} |A_{3}^* B_{10} v v_d v_{\Phi_e}^* | \frac{m_e}{M^6}\, ,
\end{align}
with a relevant physical phase $\delta$.
The experimental upper limit can be satisfied for 
\begin{align}
    M\gtrsim 100 \, \rm GeV\, .
\end{align}
Compared to the case of $n=2$, the lower bound on the mass scale $M$ becomes weaker.
However, as described in Sec.~\ref{sec:review}, the mass of $\Phi_e$ in the loop is typically $\mathcal{O}(1) \, \rm GeV$
in the case of $n=3$.
Hence, it is likely that some fine-tuning in CP-violating phases is required to satisfy the EDM constraint.
A more precise numerical estimate for the amount of tuning is left for a future study.

In addition to the one-loop contributions, a two-loop diagram with a fewer number of small VEV insertions may lead to a
sizable contribution to the electron EDM.
Fig.~\ref{sub3} describes a two-loop Bar-Zee-type diagram with inner $W$-loop.
The diagram involves the CP-violating interaction $A_3 H_e H^\dag \Phi_e^*$,
and its contribution to the electron EDM is estimated as~\cite{Barr:1990vd}
(the general formula for a two-loop Bar-Zee-type diagram has been presented in Ref.~\cite{Nakai:2016atk} (see also Ref.~\cite{Cesarotti:2018huy}))
\begin{align} 
    \frac{d_e}{e} \sim \frac{\alpha_e \zeta}{(4\pi)^3}\sqrt{2}G_F &m_e\left(3f\left(\frac{m_W^2}{m_H^2}\right)+5g\left(\frac{m_W^2}{m_H^2}\right)\right)\notag\\[1ex]
    &\times\left|\frac{A_3 v_{\Phi_e}^*}{2\sqrt{2} M^2}\right|
    \left(\sin^2\beta \tan\beta\right)\, ,
\end{align}
where $\alpha_e$ and $G_F$ denote the fine-structure constant and Fermi coupling constant, respectively,
$\zeta$ is the relevant physical phase and
\begin{align}
    f(z) &\equiv \frac{1}{2}z\int_0^1 dx\frac{1-2x(1-x)}{x(1-x)-z}\ln\left(\frac{x(1-x)}{z}\right)\, ,\notag\\
    g(z) &\equiv \frac{1}{2}z\int_0^1 dx\frac{1}{x(1-x)-z}\ln\left(\frac{x(1-x)}{z}\right)\, \notag .
\end{align}
The parameter $\beta$ is defined as
\begin{align}
    \tan\beta \equiv \frac{v}{v_e}
    \sim 10^4\, .
\end{align}
The experimental constraint can be satisfied for
\begin{align}
    M\gtrsim 8 \, \rm TeV\, ,
\end{align}
which gives a stronger bound (independent of the value of $Q_e$),
compared to those of the one-loop contributions.
Fine-tuning in CP-violating phases and/or some mechanism to suppress the phases are necessary.

In addition to the electron EDM constraint, we are able to consider
those of the neutron EDM~\cite{Baker:2006ts} and mercury EDM~\cite{Graner_2016}
where the chromo-EDM of the down quark and the electron EDM give main contributions.
The resulting constraints are found to be weaker than the one directly given by the electron EDM measurement.

\section{Loop induced B-decay}
\label{sec:loop}

\begin{figure*}
    \centering
    \includegraphics[width=0.85\textwidth]{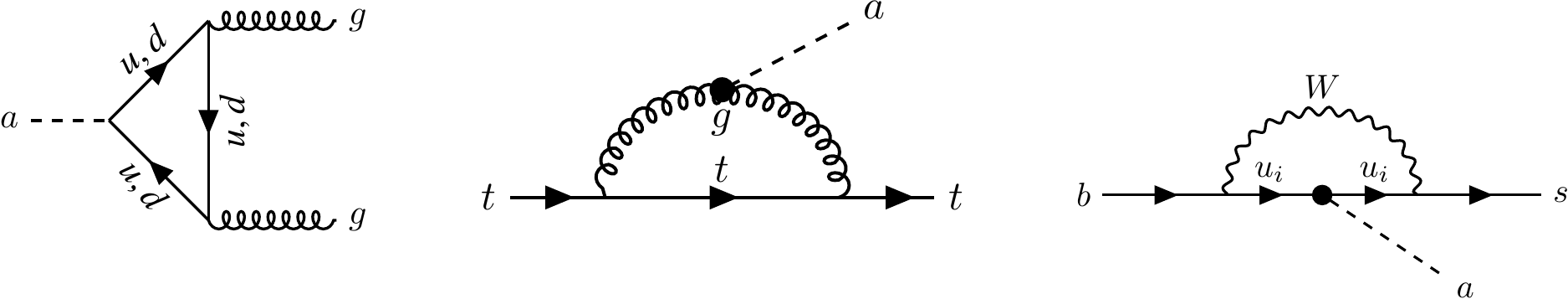}
    \caption{Feynman diagrams that generate the axion couplings to gluons and the top quark, mediating the $B$-meson decay.}
    \label{Bfig}
\end{figure*}

To evade stringent constraints from rare decay modes of heavy mesons such as $B$, $\Upsilon$ mediated by the QCD axion variant under consideration,
it has been assumed that the axion only couples to the first-generation quarks $u$, $d$, and to the electron at the tree level.
However, from a UV completion perspective, the PQ symmetry breaking, together with the electroweak symmetry breaking,
has to generate the observed CKM mixing matrix.
Therefore, inevitably, couplings of the axion to heavier generation quarks will be generated via mixing with the SM Higgs.
Progress towards understanding this has been made in Ref.~\cite{Liu:2021wap}, where a hierarchy between the VEVs
of additional Higgses and the SM Higgs has been deemed to suppress the axion couplings to the 2nd and 3rd generation quarks to
$\sim {\cal{O}}(10^{-7})$ level.

Here we point out that a gluon loop will also induce the axion couplings to the heavier quarks. Namely,
\begin{align}
    \nonumber
    {\cal L}_{aFF} &= Q_{ F, \rm eff}^{\rm PQ} \frac{m_F}{v_a} a \bar{F} i \gamma_5 F \ ,  \\  
    Q_{F, \rm  eff}^{\rm PQ} & \simeq C_F \left(\frac{\alpha_s}{4\pi}\right)^2 \ , 
    \label{Qeff}
\end{align}
where $F$ can be any 2nd or 3rd generation quark, and $C_F=4/3$ is the quadratic Casimir,
and $\alpha_s$ is the strong coupling constant.
The two-loop suppression effect in Eq.~\eqref{Qeff} comes from the fact that
the axion-gluon coupling inherits a loop factor,
generated through the up and down quarks flowing in the triangle anomaly diagrams.
Furthermore, the anomalous dimension of this coupling is $ -6 C_F (\alpha_s/4\pi)^2$,
and it is not changed significantly upon running.
Therefore, the 1-loop gluon-induced coupling to the bottom quark is $Q_{b, \rm eff}^{\rm PQ} \simeq 3 \times 10^{-4}$,
where we have used the running coupling $\alpha_s$ relevant for the process $\Upsilon \to \gamma a$,
which requires $Q_{b, \rm eff}^{\rm PQ} \lesssim 0.8 \times 10^{-2}$~\cite{Alves:2017avw}.
Hence, the 1-loop gluon-generated coupling to the $b$ quark is consistent with bounds from rare $\Upsilon$ decays.

Let us now consider the loop-mediated decays of the $B$-meson such as $B \to K a$.
This is obtained by integrating out the $W$-loop as in Fig.~\ref{Bfig}.
The effective $b$-$s$-$a$ vertex is denoted as \begin{equation}
 {\cal L}_{bsa} = -i g_{bsa} \bar{s}_L b_R a + {\rm h.c.} \ ,
 \label{Lbsa}
\end{equation} 
where we utilized the chirality-flipping nature of the diagram to infer that
the coefficient of $\bar s_{R} b_L$ is suppressed as $m_s/m_b$. The leading-log result is well-known~\cite{Batell:2009jf},
\begin{equation}
    g_{bsa} = \frac{ G_F m_W^2}{4 \sqrt{2} \pi^2} \frac{m_b}{v_a} \sum_{u_i=u,c,t} Q_{u_i, {\rm{eff}}}^{\rm PQ} \frac{m_{u_i}^2}{m_W^2} V_{u_i s}^* V_{u_i b} \ln{\left(\frac{\Lambda_{\rm UV}^2}{m_{u_i}^2} \right)} \ ,
    \label{gbsa_eqn}
\end{equation}
where we have included a summation over all the up-type quarks,
with an effective coupling to the higher generation understood as given in Eq.~\eqref{Qeff},
and $\Lambda_{\rm UV}$ is taken as the UV cut-off of the theory.
Although the top-axion coupling is loop generated, it contributes dominantly to $g_{bsa}$ as evident in Eq.~\eqref{gbsa_eqn}.
To be consistent with the beam dump decay experiment, the axion should decay into an $e^+ e^-$ pair
with lifetime $\tau_a \lesssim 10^{-13}$ s.
Hence, we employ the experimental measurement on the branching ratio of the rare decay mode $B^0 \to K^{*0} e^+ e^-$
by LHCb~\cite{LHCb:2013pra, LHCb:2015ycz},\footnote{
Here the errors are respectively statistical, systematic, and relating to the uncertainties on the $B^0 \to J/\Psi K^{*0}$
and $J/\Psi \to e^+ e^-$ branching fractions, which were used as normalization channels.}
\begin{equation}
    {\cal B}(B^0 \to K^{*0} e^+ e^-) = \left(3.1^{+0.9+0.2}_{-0.8-0.3} \pm 0.2 ({\cal B}) \right) \times 10^{-7}   \ .
    \label{LHCb_observed}
\end{equation}
Note that this measurement is only applicable for dilepton invariant mass in the range of $30$-$1000$ MeV,
which is slightly higher than the axion mass we are interested in, however,
it should be noted that the theoretical expectation from the SM is not very sensitive to the dilepton invariant mass,
and can be predicted up to the order one logarithmic correction factor as~\cite{Jager:2012uw}
\begin{equation}
    {\cal B}^{\rm SM}(B^0 \to K^{*0} e^+ e^-) \simeq 2.43^{+0.66}_{-0.47} \times 10^{-7} \ .
    \label{BtoKee_SM}
\end{equation}
Given the axion coupling in Eq.~\eqref{Lbsa}, the BSM contribution can be written as 
\begin{align}
    {\cal B}^{\rm BSM}&(B^0 \to K^{*0} e^+ e^-) \nonumber \\
    &= \frac{|g_{bsa}|^2}{16\pi m_B \Gamma_B} \lambda_{B, K^* a}^{1/2} |\langle K^*| \bar{s}_L b_R|B \rangle|^2 \ ,
\end{align}
where 
\begin{align}
    \nonumber
   & \lambda_{x, yz}  \equiv \left[1-\left(\frac{m_y}{m_x}-\frac{m_z}{m_x}\right)^2\right] \left[1-\left(\frac{m_y}{m_x}+\frac{m_z}{m_x}\right)^2\right] \ , \nonumber \\[1ex] 
   & |\langle K^*| \bar{s}_L b_R|B \rangle|^2 = \frac{1}{4} \frac{m_B^4 \lambda_{B, K^* a}}{(m_b+m_s)^2} [{\cal A}_{K^*}(m_a^2)]^2 \ , \\[1ex] 
   & {\cal A}_{K^*}(q^2) = \frac{1.36}{1-q^2/27.9 \, {\rm GeV}^2}-\frac{0.99}{1-q^2/36.8 \, {\rm GeV}^2} \ . \nonumber
\end{align}
The pion-phobic QCD axion variant, which by an ad-hoc model construction only couples to
the first-generation quarks at the tree level,
when considered together with the gluon loop generated coupling to the top quark,
can mediate the $B$-decay with a contribution much larger than that of the SM.
For numerical illustration,
taking $m_a = 10$ MeV, we find
\begin{equation}
    \frac{{\cal B}^{\rm BSM}(B^0 \to K^{*0}a (\to e^+ e^-))}{{\cal B}^{\rm SM}(B^0 \to K^{*0} e^+ e^-)}  \simeq 10^{3} \ln{\left(\frac{\Lambda_{\rm UV}}{m_{t}} \right)}^2 \ ,
\end{equation}
where we have used the QCD relation in Eq.~\eqref{Eq:QCD_relation} together with $v_a = N_{\rm DW} f_a$ for this model,
which fixes $v_a \simeq 1.7$ GeV for $m_a = 10$ MeV.
This BSM contribution to the rare meson decay increases with increasing $m_a$, as the corresponding $v_a$ decreases. Hence, in light of the LHCb analysis~\cite{LHCb:2013pra, LHCb:2015ycz},
the pion-phobic QCD axion variant is ruled out for $m_a > 30$ MeV mass range.

The current most precise measurement of the branching ratio for $B \to K^{(*)} e^+ e^-$ comes from
Belle~\cite{Belle:2009zue} and BaBar~\cite{BaBar:2008jdv} collaborations.
However, they used a veto of $q^2 \lesssim 0.1$ GeV$^2/c^4$ to reduce the background.
Therefore, we cannot use their result to constrain the axion mass below $\sim$ 300 MeV.
We note that below 30 MeV dilepton invariant mass,
multiple scatterings obscure the angular analysis of the outgoing electrons in $B^0 \to K^{*0} e^+ e^-$, however, if one is only interested in searching for this mode without the angular correlation, the background contamination only comes from the photon converting to $e^+ e^-$ in the mode $B^0 \to K^{*0} \gamma$. It is also conceivable that meaningful constraints can be obtained by analyzing the off-shell contribution of this axion variant to the lower dilepton mass bins. Hence, there is a strong motivation for the LHCb and Belle II experiments to make an updated analysis and push the search for this mode towards the lower end of the dilepton invariant mass from 30 MeV down to a few MeV that will determine the fate of the model conclusively.

\section{Conclusions
\label{sec:discussion}}

We have considered new theoretical and phenomenological constraints on the ${\cal O}$(10) MeV
pion-phobic QCD axion variant that has been argued to be still viable~\cite{Alves:2017avw}.
Due to the very low PQ breaking scale,
the phase transition takes place after inflation, and the string-domain wall network is formed.
We have investigated the interplay of the lifetime of the network
and the quality issue of the PQ symmetry due to a necessary bias term.
It was found that an appropriate choice of a higher-dimensional operator by some model-building can 
solve the domain wall problem with the high-quality of the PQ symmetry.
From a UV completion perspective,
as the PQ charges of the up and down quarks and electron have to be different in the model,
one has to introduce multiple new scalar fields.
This brings the model a plethora of new CP-violating terms that manifest themselves as EDMs.
We have identified the dominant electron EDM contribution and analyzed the resulting constraints.
Finally, we have pointed out that the gluon loop-induced top-quark coupling of the axion contributes to rare $B$-decay modes,
and in particular to $B \to K^{(*)} a( \to e^+ e^-)$ for the current axion model.
Experimental data for this mode already exclude the model for the axion mass larger than 30 MeV.
There is a strong motivation for pushing the experimental analysis to a lower dilepton mass window
by the LHCb and Belle collaborations, which will dictate the fate of the model.

\section*{Acknowledgments}
We thank Chia-Wei Liu and  Yoshihiro Shigekami
for useful discussions.


\bibliography{reference}

\begin{thebibliography}{72}%
\makeatletter
\providecommand \@ifxundefined [1]{%
 \@ifx{#1\undefined}
}%
\providecommand \@ifnum [1]{%
 \ifnum #1\expandafter \@firstoftwo
 \else \expandafter \@secondoftwo
 \fi
}%
\providecommand \@ifx [1]{%
 \ifx #1\expandafter \@firstoftwo
 \else \expandafter \@secondoftwo
 \fi
}%
\providecommand \natexlab [1]{#1}%
\providecommand \enquote  [1]{``#1''}%
\providecommand \bibnamefont  [1]{#1}%
\providecommand \bibfnamefont [1]{#1}%
\providecommand \citenamefont [1]{#1}%
\providecommand \href@noop [0]{\@secondoftwo}%
\providecommand \href [0]{\begingroup \@sanitize@url \@href}%
\providecommand \@href[1]{\@@startlink{#1}\@@href}%
\providecommand \@@href[1]{\endgroup#1\@@endlink}%
\providecommand \@sanitize@url [0]{\catcode `\\12\catcode `\$12\catcode
  `\&12\catcode `\#12\catcode `\^12\catcode `\_12\catcode `\%12\relax}%
\providecommand \@@startlink[1]{}%
\providecommand \@@endlink[0]{}%
\providecommand \url  [0]{\begingroup\@sanitize@url \@url }%
\providecommand \@url [1]{\endgroup\@href {#1}{\urlprefix }}%
\providecommand \urlprefix  [0]{URL }%
\providecommand \Eprint [0]{\href }%
\providecommand \doibase [0]{http://dx.doi.org/}%
\providecommand \selectlanguage [0]{\@gobble}%
\providecommand \bibinfo  [0]{\@secondoftwo}%
\providecommand \bibfield  [0]{\@secondoftwo}%
\providecommand \translation [1]{[#1]}%
\providecommand \BibitemOpen [0]{}%
\providecommand \bibitemStop [0]{}%
\providecommand \bibitemNoStop [0]{.\EOS\space}%
\providecommand \EOS [0]{\spacefactor3000\relax}%
\providecommand \BibitemShut  [1]{\csname bibitem#1\endcsname}%
\let\auto@bib@innerbib\@empty
\bibitem [{\citenamefont {Ellis}\ and\ \citenamefont
  {Gaillard}(1979)}]{Ellis:1978hq}%
  \BibitemOpen
  \bibfield  {author} {\bibinfo {author} {\bibfnamefont {J.~R.}\ \bibnamefont
  {Ellis}}\ and\ \bibinfo {author} {\bibfnamefont {M.~K.}\ \bibnamefont
  {Gaillard}},\ }\href {\doibase 10.1016/0550-3213(79)90297-9} {\bibfield
  {journal} {\bibinfo  {journal} {Nucl. Phys. B}\ }\textbf {\bibinfo {volume}
  {150}},\ \bibinfo {pages} {141} (\bibinfo {year} {1979})}\BibitemShut
  {NoStop}%
\bibitem [{\citenamefont {Workman}\ \emph {et~al.}(2022)\citenamefont {Workman}
  \emph {et~al.}}]{ParticleDataGroup:2022pth}%
  \BibitemOpen
  \bibfield  {author} {\bibinfo {author} {\bibfnamefont {R.~L.}\ \bibnamefont
  {Workman}} \emph {et~al.} (\bibinfo {collaboration} {Particle Data Group}),\
  }\href {\doibase 10.1093/ptep/ptac097} {\bibfield  {journal} {\bibinfo
  {journal} {PTEP}\ }\textbf {\bibinfo {volume} {2022}},\ \bibinfo {pages}
  {083C01} (\bibinfo {year} {2022})}\BibitemShut {NoStop}%
\bibitem [{\citenamefont {Alexandrou}\ \emph {et~al.}(2020)\citenamefont
  {Alexandrou}, \citenamefont {Finkenrath}, \citenamefont {Funcke},
  \citenamefont {Jansen}, \citenamefont {Kostrzewa}, \citenamefont {Pittler},\
  and\ \citenamefont {Urbach}}]{Alexandrou:2020bkd}%
  \BibitemOpen
  \bibfield  {author} {\bibinfo {author} {\bibfnamefont {C.}~\bibnamefont
  {Alexandrou}}, \bibinfo {author} {\bibfnamefont {J.}~\bibnamefont
  {Finkenrath}}, \bibinfo {author} {\bibfnamefont {L.}~\bibnamefont {Funcke}},
  \bibinfo {author} {\bibfnamefont {K.}~\bibnamefont {Jansen}}, \bibinfo
  {author} {\bibfnamefont {B.}~\bibnamefont {Kostrzewa}}, \bibinfo {author}
  {\bibfnamefont {F.}~\bibnamefont {Pittler}}, \ and\ \bibinfo {author}
  {\bibfnamefont {C.}~\bibnamefont {Urbach}},\ }\href {\doibase
  10.1103/PhysRevLett.125.232001} {\bibfield  {journal} {\bibinfo  {journal}
  {Phys. Rev. Lett.}\ }\textbf {\bibinfo {volume} {125}},\ \bibinfo {pages}
  {232001} (\bibinfo {year} {2020})},\ \Eprint
  {http://arxiv.org/abs/2002.07802} {arXiv:2002.07802 [hep-lat]} \BibitemShut
  {NoStop}%
\bibitem [{\citenamefont {Peccei}\ and\ \citenamefont
  {Quinn}(1977)}]{Peccei:1977hh}%
  \BibitemOpen
  \bibfield  {author} {\bibinfo {author} {\bibfnamefont {R.~D.}\ \bibnamefont
  {Peccei}}\ and\ \bibinfo {author} {\bibfnamefont {H.~R.}\ \bibnamefont
  {Quinn}},\ }\href {\doibase 10.1103/PhysRevLett.38.1440} {\bibfield
  {journal} {\bibinfo  {journal} {Phys. Rev. Lett.}\ }\textbf {\bibinfo
  {volume} {38}},\ \bibinfo {pages} {1440} (\bibinfo {year}
  {1977})}\BibitemShut {NoStop}%
\bibitem [{\citenamefont {Weinberg}(1978)}]{Weinberg:1977ma}%
  \BibitemOpen
  \bibfield  {author} {\bibinfo {author} {\bibfnamefont {S.}~\bibnamefont
  {Weinberg}},\ }\href {\doibase 10.1103/PhysRevLett.40.223} {\bibfield
  {journal} {\bibinfo  {journal} {Phys. Rev. Lett.}\ }\textbf {\bibinfo
  {volume} {40}},\ \bibinfo {pages} {223} (\bibinfo {year} {1978})}\BibitemShut
  {NoStop}%
\bibitem [{\citenamefont {Wilczek}(1978)}]{Wilczek:1977pj}%
  \BibitemOpen
  \bibfield  {author} {\bibinfo {author} {\bibfnamefont {F.}~\bibnamefont
  {Wilczek}},\ }\href {\doibase 10.1103/PhysRevLett.40.279} {\bibfield
  {journal} {\bibinfo  {journal} {Phys. Rev. Lett.}\ }\textbf {\bibinfo
  {volume} {40}},\ \bibinfo {pages} {279} (\bibinfo {year} {1978})}\BibitemShut
  {NoStop}%
\bibitem [{\citenamefont {Kim}(1979)}]{Kim:1979if}%
  \BibitemOpen
  \bibfield  {author} {\bibinfo {author} {\bibfnamefont {J.~E.}\ \bibnamefont
  {Kim}},\ }\href {\doibase 10.1103/PhysRevLett.43.103} {\bibfield  {journal}
  {\bibinfo  {journal} {Phys. Rev. Lett.}\ }\textbf {\bibinfo {volume} {43}},\
  \bibinfo {pages} {103} (\bibinfo {year} {1979})}\BibitemShut {NoStop}%
\bibitem [{\citenamefont {Dine}\ \emph {et~al.}(1981)\citenamefont {Dine},
  \citenamefont {Fischler},\ and\ \citenamefont {Srednicki}}]{Dine:1981rt}%
  \BibitemOpen
  \bibfield  {author} {\bibinfo {author} {\bibfnamefont {M.}~\bibnamefont
  {Dine}}, \bibinfo {author} {\bibfnamefont {W.}~\bibnamefont {Fischler}}, \
  and\ \bibinfo {author} {\bibfnamefont {M.}~\bibnamefont {Srednicki}},\ }\href
  {\doibase 10.1016/0370-2693(81)90590-6} {\bibfield  {journal} {\bibinfo
  {journal} {Phys. Lett. B}\ }\textbf {\bibinfo {volume} {104}},\ \bibinfo
  {pages} {199} (\bibinfo {year} {1981})}\BibitemShut {NoStop}%
\bibitem [{\citenamefont {Alves}\ and\ \citenamefont
  {Weiner}(2018)}]{Alves:2017avw}%
  \BibitemOpen
  \bibfield  {author} {\bibinfo {author} {\bibfnamefont {D.~S.~M.}\
  \bibnamefont {Alves}}\ and\ \bibinfo {author} {\bibfnamefont
  {N.}~\bibnamefont {Weiner}},\ }\href {\doibase 10.1007/JHEP07(2018)092}
  {\bibfield  {journal} {\bibinfo  {journal} {JHEP}\ }\textbf {\bibinfo
  {volume} {07}},\ \bibinfo {pages} {092} (\bibinfo {year} {2018})},\ \Eprint
  {http://arxiv.org/abs/1710.03764} {arXiv:1710.03764 [hep-ph]} \BibitemShut
  {NoStop}%
\bibitem [{\citenamefont {Krasznahorkay}\ \emph {et~al.}(2016)\citenamefont
  {Krasznahorkay} \emph {et~al.}}]{Krasznahorkay:2015iga}%
  \BibitemOpen
  \bibfield  {author} {\bibinfo {author} {\bibfnamefont {A.~J.}\ \bibnamefont
  {Krasznahorkay}} \emph {et~al.},\ }\href {\doibase
  10.1103/PhysRevLett.116.042501} {\bibfield  {journal} {\bibinfo  {journal}
  {Phys. Rev. Lett.}\ }\textbf {\bibinfo {volume} {116}},\ \bibinfo {pages}
  {042501} (\bibinfo {year} {2016})},\ \Eprint
  {http://arxiv.org/abs/1504.01527} {arXiv:1504.01527 [nucl-ex]} \BibitemShut
  {NoStop}%
\bibitem [{\citenamefont {Liu}\ \emph {et~al.}(2021)\citenamefont {Liu},
  \citenamefont {McGinnis}, \citenamefont {Wagner},\ and\ \citenamefont
  {Wang}}]{Liu:2021wap}%
  \BibitemOpen
  \bibfield  {author} {\bibinfo {author} {\bibfnamefont {J.}~\bibnamefont
  {Liu}}, \bibinfo {author} {\bibfnamefont {N.}~\bibnamefont {McGinnis}},
  \bibinfo {author} {\bibfnamefont {C.~E.~M.}\ \bibnamefont {Wagner}}, \ and\
  \bibinfo {author} {\bibfnamefont {X.-P.}\ \bibnamefont {Wang}},\ }\href
  {\doibase 10.1007/JHEP05(2021)138} {\bibfield  {journal} {\bibinfo  {journal}
  {JHEP}\ }\textbf {\bibinfo {volume} {05}},\ \bibinfo {pages} {138} (\bibinfo
  {year} {2021})},\ \Eprint {http://arxiv.org/abs/2102.10118} {arXiv:2102.10118
  [hep-ph]} \BibitemShut {NoStop}%
\bibitem [{\citenamefont {Krauss}\ and\ \citenamefont
  {Nash}(1988)}]{Krauss:1987ud}%
  \BibitemOpen
  \bibfield  {author} {\bibinfo {author} {\bibfnamefont {L.~M.}\ \bibnamefont
  {Krauss}}\ and\ \bibinfo {author} {\bibfnamefont {D.~J.}\ \bibnamefont
  {Nash}},\ }\href {\doibase 10.1016/0370-2693(88)91864-3} {\bibfield
  {journal} {\bibinfo  {journal} {Phys. Lett. B}\ }\textbf {\bibinfo {volume}
  {202}},\ \bibinfo {pages} {560} (\bibinfo {year} {1988})}\BibitemShut
  {NoStop}%
\bibitem [{\citenamefont {Alves}(2021)}]{Alves:2020xhf}%
  \BibitemOpen
  \bibfield  {author} {\bibinfo {author} {\bibfnamefont {D.~S.~M.}\
  \bibnamefont {Alves}},\ }\href {\doibase 10.1103/PhysRevD.103.055018}
  {\bibfield  {journal} {\bibinfo  {journal} {Phys. Rev. D}\ }\textbf {\bibinfo
  {volume} {103}},\ \bibinfo {pages} {055018} (\bibinfo {year} {2021})},\
  \Eprint {http://arxiv.org/abs/2009.05578} {arXiv:2009.05578 [hep-ph]}
  \BibitemShut {NoStop}%
\bibitem [{\citenamefont {Hostert}\ and\ \citenamefont
  {Pospelov}(2022)}]{Hostert:2020xku}%
  \BibitemOpen
  \bibfield  {author} {\bibinfo {author} {\bibfnamefont {M.}~\bibnamefont
  {Hostert}}\ and\ \bibinfo {author} {\bibfnamefont {M.}~\bibnamefont
  {Pospelov}},\ }\href {\doibase 10.1103/PhysRevD.105.015017} {\bibfield
  {journal} {\bibinfo  {journal} {Phys. Rev. D}\ }\textbf {\bibinfo {volume}
  {105}},\ \bibinfo {pages} {015017} (\bibinfo {year} {2022})},\ \Eprint
  {http://arxiv.org/abs/2012.02142} {arXiv:2012.02142 [hep-ph]} \BibitemShut
  {NoStop}%
\bibitem [{\citenamefont {Cortina~Gil}\ \emph {et~al.}(2023)\citenamefont
  {Cortina~Gil} \emph {et~al.}}]{NA62:2023rvm}%
  \BibitemOpen
  \bibfield  {author} {\bibinfo {author} {\bibfnamefont {E.}~\bibnamefont
  {Cortina~Gil}} \emph {et~al.} (\bibinfo {collaboration} {NA62}),\ }\href
  {\doibase 10.1016/j.physletb.2023.138193} {\bibfield  {journal} {\bibinfo
  {journal} {Phys. Lett. B}\ }\textbf {\bibinfo {volume} {846}},\ \bibinfo
  {pages} {138193} (\bibinfo {year} {2023})},\ \Eprint
  {http://arxiv.org/abs/2307.04579} {arXiv:2307.04579 [hep-ex]} \BibitemShut
  {NoStop}%
\bibitem [{\citenamefont {Aaij}\ \emph {et~al.}(2013)\citenamefont {Aaij} \emph
  {et~al.}}]{LHCb:2013pra}%
  \BibitemOpen
  \bibfield  {author} {\bibinfo {author} {\bibfnamefont {R.}~\bibnamefont
  {Aaij}} \emph {et~al.} (\bibinfo {collaboration} {LHCb}),\ }\href {\doibase
  10.1007/JHEP05(2013)159} {\bibfield  {journal} {\bibinfo  {journal} {JHEP}\
  }\textbf {\bibinfo {volume} {05}},\ \bibinfo {pages} {159} (\bibinfo {year}
  {2013})},\ \Eprint {http://arxiv.org/abs/1304.3035} {arXiv:1304.3035
  [hep-ex]} \BibitemShut {NoStop}%
\bibitem [{\citenamefont {Donnelly}\ \emph {et~al.}(1978)\citenamefont
  {Donnelly}, \citenamefont {Freedman}, \citenamefont {Lytel}, \citenamefont
  {Peccei},\ and\ \citenamefont {Schwartz}}]{Donnelly:1978ty}%
  \BibitemOpen
  \bibfield  {author} {\bibinfo {author} {\bibfnamefont {T.~W.}\ \bibnamefont
  {Donnelly}}, \bibinfo {author} {\bibfnamefont {S.~J.}\ \bibnamefont
  {Freedman}}, \bibinfo {author} {\bibfnamefont {R.~S.}\ \bibnamefont {Lytel}},
  \bibinfo {author} {\bibfnamefont {R.~D.}\ \bibnamefont {Peccei}}, \ and\
  \bibinfo {author} {\bibfnamefont {M.}~\bibnamefont {Schwartz}},\ }\href
  {\doibase 10.1103/PhysRevD.18.1607} {\bibfield  {journal} {\bibinfo
  {journal} {Phys. Rev. D}\ }\textbf {\bibinfo {volume} {18}},\ \bibinfo
  {pages} {1607} (\bibinfo {year} {1978})}\BibitemShut {NoStop}%
\bibitem [{\citenamefont {Wilczek}(1977)}]{Wilczek:1977zn}%
  \BibitemOpen
  \bibfield  {author} {\bibinfo {author} {\bibfnamefont {F.}~\bibnamefont
  {Wilczek}},\ }\href {\doibase 10.1103/PhysRevLett.39.1304} {\bibfield
  {journal} {\bibinfo  {journal} {Phys. Rev. Lett.}\ }\textbf {\bibinfo
  {volume} {39}},\ \bibinfo {pages} {1304} (\bibinfo {year}
  {1977})}\BibitemShut {NoStop}%
\bibitem [{\citenamefont {Hall}\ and\ \citenamefont
  {Wise}(1981)}]{Hall:1981bc}%
  \BibitemOpen
  \bibfield  {author} {\bibinfo {author} {\bibfnamefont {L.~J.}\ \bibnamefont
  {Hall}}\ and\ \bibinfo {author} {\bibfnamefont {M.~B.}\ \bibnamefont
  {Wise}},\ }\href {\doibase 10.1016/0550-3213(81)90469-7} {\bibfield
  {journal} {\bibinfo  {journal} {Nucl. Phys. B}\ }\textbf {\bibinfo {volume}
  {187}},\ \bibinfo {pages} {397} (\bibinfo {year} {1981})}\BibitemShut
  {NoStop}%
\bibitem [{\citenamefont {Albrecht}\ \emph {et~al.}(1986)\citenamefont
  {Albrecht} \emph {et~al.}}]{ARGUS:1986ytx}%
  \BibitemOpen
  \bibfield  {author} {\bibinfo {author} {\bibfnamefont {H.}~\bibnamefont
  {Albrecht}} \emph {et~al.} (\bibinfo {collaboration} {ARGUS}),\ }\href
  {\doibase 10.1016/0370-2693(86)90501-0} {\bibfield  {journal} {\bibinfo
  {journal} {Phys. Lett. B}\ }\textbf {\bibinfo {volume} {179}},\ \bibinfo
  {pages} {403} (\bibinfo {year} {1986})}\BibitemShut {NoStop}%
\bibitem [{\citenamefont {Hsueh}\ and\ \citenamefont
  {Palestini}(1992)}]{Hsueh:1992ex}%
  \BibitemOpen
  \bibfield  {author} {\bibinfo {author} {\bibfnamefont {S.~Y.}\ \bibnamefont
  {Hsueh}}\ and\ \bibinfo {author} {\bibfnamefont {S.}~\bibnamefont
  {Palestini}},\ }\href {\doibase 10.1103/PhysRevD.45.R2181} {\bibfield
  {journal} {\bibinfo  {journal} {Phys. Rev. D}\ }\textbf {\bibinfo {volume}
  {45}},\ \bibinfo {pages} {R2181} (\bibinfo {year} {1992})}\BibitemShut
  {NoStop}%
\bibitem [{\citenamefont {Armstrong}\ \emph {et~al.}(1993)\citenamefont
  {Armstrong} \emph {et~al.}}]{E760:1992pgt}%
  \BibitemOpen
  \bibfield  {author} {\bibinfo {author} {\bibfnamefont {T.~A.}\ \bibnamefont
  {Armstrong}} \emph {et~al.} (\bibinfo {collaboration} {E760}),\ }\href
  {\doibase 10.1103/PhysRevD.47.772} {\bibfield  {journal} {\bibinfo  {journal}
  {Phys. Rev. D}\ }\textbf {\bibinfo {volume} {47}},\ \bibinfo {pages} {772}
  (\bibinfo {year} {1993})}\BibitemShut {NoStop}%
\bibitem [{\citenamefont {Bai}\ \emph {et~al.}(1995)\citenamefont {Bai} \emph
  {et~al.}}]{BES:1995wyo}%
  \BibitemOpen
  \bibfield  {author} {\bibinfo {author} {\bibfnamefont {J.~Z.}\ \bibnamefont
  {Bai}} \emph {et~al.} (\bibinfo {collaboration} {BES}),\ }\href {\doibase
  10.1016/0370-2693(95)00712-T} {\bibfield  {journal} {\bibinfo  {journal}
  {Phys. Lett. B}\ }\textbf {\bibinfo {volume} {355}},\ \bibinfo {pages} {374}
  (\bibinfo {year} {1995})},\ \bibinfo {note} {[Erratum: Phys.Lett.B 363, 267
  (1995)]}\BibitemShut {NoStop}%
\bibitem [{\citenamefont {Adams}\ \emph {et~al.}(2006)\citenamefont {Adams}
  \emph {et~al.}}]{CLEO:2005cdx}%
  \BibitemOpen
  \bibfield  {author} {\bibinfo {author} {\bibfnamefont {G.~S.}\ \bibnamefont
  {Adams}} \emph {et~al.} (\bibinfo {collaboration} {CLEO}),\ }\href {\doibase
  10.1103/PhysRevD.73.051103} {\bibfield  {journal} {\bibinfo  {journal} {Phys.
  Rev. D}\ }\textbf {\bibinfo {volume} {73}},\ \bibinfo {pages} {051103}
  (\bibinfo {year} {2006})},\ \Eprint {http://arxiv.org/abs/hep-ex/0512046}
  {arXiv:hep-ex/0512046} \BibitemShut {NoStop}%
\bibitem [{\citenamefont {Bergsma}\ \emph {et~al.}(1985)\citenamefont {Bergsma}
  \emph {et~al.}}]{CHARM:1985anb}%
  \BibitemOpen
  \bibfield  {author} {\bibinfo {author} {\bibfnamefont {F.}~\bibnamefont
  {Bergsma}} \emph {et~al.} (\bibinfo {collaboration} {CHARM}),\ }\href
  {\doibase 10.1016/0370-2693(85)90400-9} {\bibfield  {journal} {\bibinfo
  {journal} {Phys. Lett. B}\ }\textbf {\bibinfo {volume} {157}},\ \bibinfo
  {pages} {458} (\bibinfo {year} {1985})}\BibitemShut {NoStop}%
\bibitem [{\citenamefont {Bjorken}\ \emph {et~al.}(1988)\citenamefont
  {Bjorken}, \citenamefont {Ecklund}, \citenamefont {Nelson}, \citenamefont
  {Abashian}, \citenamefont {Church}, \citenamefont {Lu}, \citenamefont {Mo},
  \citenamefont {Nunamaker},\ and\ \citenamefont {Rassmann}}]{Bjorken:1988as}%
  \BibitemOpen
  \bibfield  {author} {\bibinfo {author} {\bibfnamefont {J.~D.}\ \bibnamefont
  {Bjorken}}, \bibinfo {author} {\bibfnamefont {S.}~\bibnamefont {Ecklund}},
  \bibinfo {author} {\bibfnamefont {W.~R.}\ \bibnamefont {Nelson}}, \bibinfo
  {author} {\bibfnamefont {A.}~\bibnamefont {Abashian}}, \bibinfo {author}
  {\bibfnamefont {C.}~\bibnamefont {Church}}, \bibinfo {author} {\bibfnamefont
  {B.}~\bibnamefont {Lu}}, \bibinfo {author} {\bibfnamefont {L.~W.}\
  \bibnamefont {Mo}}, \bibinfo {author} {\bibfnamefont {T.~A.}\ \bibnamefont
  {Nunamaker}}, \ and\ \bibinfo {author} {\bibfnamefont {P.}~\bibnamefont
  {Rassmann}},\ }\href {\doibase 10.1103/PhysRevD.38.3375} {\bibfield
  {journal} {\bibinfo  {journal} {Phys. Rev. D}\ }\textbf {\bibinfo {volume}
  {38}},\ \bibinfo {pages} {3375} (\bibinfo {year} {1988})}\BibitemShut
  {NoStop}%
\bibitem [{\citenamefont {Davier}\ and\ \citenamefont
  {Nguyen~Ngoc}(1989)}]{Davier:1989wz}%
  \BibitemOpen
  \bibfield  {author} {\bibinfo {author} {\bibfnamefont {M.}~\bibnamefont
  {Davier}}\ and\ \bibinfo {author} {\bibfnamefont {H.}~\bibnamefont
  {Nguyen~Ngoc}},\ }\href {\doibase 10.1016/0370-2693(89)90174-3} {\bibfield
  {journal} {\bibinfo  {journal} {Phys. Lett. B}\ }\textbf {\bibinfo {volume}
  {229}},\ \bibinfo {pages} {150} (\bibinfo {year} {1989})}\BibitemShut
  {NoStop}%
\bibitem [{\citenamefont {Blumlein}\ \emph {et~al.}(1991)\citenamefont
  {Blumlein} \emph {et~al.}}]{Blumlein:1990ay}%
  \BibitemOpen
  \bibfield  {author} {\bibinfo {author} {\bibfnamefont {J.}~\bibnamefont
  {Blumlein}} \emph {et~al.},\ }\href {\doibase 10.1007/BF01548556} {\bibfield
  {journal} {\bibinfo  {journal} {Z. Phys. C}\ }\textbf {\bibinfo {volume}
  {51}},\ \bibinfo {pages} {341} (\bibinfo {year} {1991})}\BibitemShut
  {NoStop}%
\bibitem [{\citenamefont {Eichler}\ \emph {et~al.}(1986)\citenamefont {Eichler}
  \emph {et~al.}}]{SINDRUM:1986klz}%
  \BibitemOpen
  \bibfield  {author} {\bibinfo {author} {\bibfnamefont {R.}~\bibnamefont
  {Eichler}} \emph {et~al.} (\bibinfo {collaboration} {SINDRUM}),\ }\href
  {\doibase 10.1016/0370-2693(86)90339-4} {\bibfield  {journal} {\bibinfo
  {journal} {Phys. Lett. B}\ }\textbf {\bibinfo {volume} {175}},\ \bibinfo
  {pages} {101} (\bibinfo {year} {1986})}\BibitemShut {NoStop}%
\bibitem [{\citenamefont {Chala}\ \emph {et~al.}(2021)\citenamefont {Chala},
  \citenamefont {Guedes}, \citenamefont {Ramos},\ and\ \citenamefont
  {Santiago}}]{Chala:2020wvs}%
  \BibitemOpen
  \bibfield  {author} {\bibinfo {author} {\bibfnamefont {M.}~\bibnamefont
  {Chala}}, \bibinfo {author} {\bibfnamefont {G.}~\bibnamefont {Guedes}},
  \bibinfo {author} {\bibfnamefont {M.}~\bibnamefont {Ramos}}, \ and\ \bibinfo
  {author} {\bibfnamefont {J.}~\bibnamefont {Santiago}},\ }\href {\doibase
  10.1140/epjc/s10052-021-08968-2} {\bibfield  {journal} {\bibinfo  {journal}
  {Eur. Phys. J. C}\ }\textbf {\bibinfo {volume} {81}},\ \bibinfo {pages} {181}
  (\bibinfo {year} {2021})},\ \Eprint {http://arxiv.org/abs/2012.09017}
  {arXiv:2012.09017 [hep-ph]} \BibitemShut {NoStop}%
\bibitem [{\citenamefont {Bauer}\ \emph {et~al.}(2021)\citenamefont {Bauer},
  \citenamefont {Neubert}, \citenamefont {Renner}, \citenamefont {Schnubel},\
  and\ \citenamefont {Thamm}}]{Bauer:2020jbp}%
  \BibitemOpen
  \bibfield  {author} {\bibinfo {author} {\bibfnamefont {M.}~\bibnamefont
  {Bauer}}, \bibinfo {author} {\bibfnamefont {M.}~\bibnamefont {Neubert}},
  \bibinfo {author} {\bibfnamefont {S.}~\bibnamefont {Renner}}, \bibinfo
  {author} {\bibfnamefont {M.}~\bibnamefont {Schnubel}}, \ and\ \bibinfo
  {author} {\bibfnamefont {A.}~\bibnamefont {Thamm}},\ }\href {\doibase
  10.1007/JHEP04(2021)063} {\bibfield  {journal} {\bibinfo  {journal} {JHEP}\
  }\textbf {\bibinfo {volume} {04}},\ \bibinfo {pages} {063} (\bibinfo {year}
  {2021})},\ \Eprint {http://arxiv.org/abs/2012.12272} {arXiv:2012.12272
  [hep-ph]} \BibitemShut {NoStop}%
\bibitem [{\citenamefont {Di~Luzio}\ \emph {et~al.}(2022)\citenamefont
  {Di~Luzio}, \citenamefont {Mescia}, \citenamefont {Nardi},\ and\
  \citenamefont {Okawa}}]{DiLuzio:2022tyc}%
  \BibitemOpen
  \bibfield  {author} {\bibinfo {author} {\bibfnamefont {L.}~\bibnamefont
  {Di~Luzio}}, \bibinfo {author} {\bibfnamefont {F.}~\bibnamefont {Mescia}},
  \bibinfo {author} {\bibfnamefont {E.}~\bibnamefont {Nardi}}, \ and\ \bibinfo
  {author} {\bibfnamefont {S.}~\bibnamefont {Okawa}},\ }\href {\doibase
  10.1103/PhysRevD.106.055016} {\bibfield  {journal} {\bibinfo  {journal}
  {Phys. Rev. D}\ }\textbf {\bibinfo {volume} {106}},\ \bibinfo {pages}
  {055016} (\bibinfo {year} {2022})},\ \Eprint
  {http://arxiv.org/abs/2205.15326} {arXiv:2205.15326 [hep-ph]} \BibitemShut
  {NoStop}%
\bibitem [{\citenamefont {Parker}\ \emph {et~al.}(2018)\citenamefont {Parker},
  \citenamefont {Yu}, \citenamefont {Zhong}, \citenamefont {Estey},\ and\
  \citenamefont {M\"uller}}]{Parker:2018vye}%
  \BibitemOpen
  \bibfield  {author} {\bibinfo {author} {\bibfnamefont {R.~H.}\ \bibnamefont
  {Parker}}, \bibinfo {author} {\bibfnamefont {C.}~\bibnamefont {Yu}}, \bibinfo
  {author} {\bibfnamefont {W.}~\bibnamefont {Zhong}}, \bibinfo {author}
  {\bibfnamefont {B.}~\bibnamefont {Estey}}, \ and\ \bibinfo {author}
  {\bibfnamefont {H.}~\bibnamefont {M\"uller}},\ }\href {\doibase
  10.1126/science.aap7706} {\bibfield  {journal} {\bibinfo  {journal}
  {Science}\ }\textbf {\bibinfo {volume} {360}},\ \bibinfo {pages} {191}
  (\bibinfo {year} {2018})},\ \Eprint {http://arxiv.org/abs/1812.04130}
  {arXiv:1812.04130 [physics.atom-ph]} \BibitemShut {NoStop}%
\bibitem [{\citenamefont {Morel}\ \emph {et~al.}(2020)\citenamefont {Morel},
  \citenamefont {Yao}, \citenamefont {Clad\'e},\ and\ \citenamefont
  {Guellati-Kh\'elifa}}]{Morel:2020dww}%
  \BibitemOpen
  \bibfield  {author} {\bibinfo {author} {\bibfnamefont {L.}~\bibnamefont
  {Morel}}, \bibinfo {author} {\bibfnamefont {Z.}~\bibnamefont {Yao}}, \bibinfo
  {author} {\bibfnamefont {P.}~\bibnamefont {Clad\'e}}, \ and\ \bibinfo
  {author} {\bibfnamefont {S.}~\bibnamefont {Guellati-Kh\'elifa}},\ }\href
  {\doibase 10.1038/s41586-020-2964-7} {\bibfield  {journal} {\bibinfo
  {journal} {Nature}\ }\textbf {\bibinfo {volume} {588}},\ \bibinfo {pages}
  {61} (\bibinfo {year} {2020})}\BibitemShut {NoStop}%
\bibitem [{\citenamefont {Fan}\ \emph {et~al.}(2023)\citenamefont {Fan},
  \citenamefont {Myers}, \citenamefont {Sukra},\ and\ \citenamefont
  {Gabrielse}}]{Fan:2022eto}%
  \BibitemOpen
  \bibfield  {author} {\bibinfo {author} {\bibfnamefont {X.}~\bibnamefont
  {Fan}}, \bibinfo {author} {\bibfnamefont {T.~G.}\ \bibnamefont {Myers}},
  \bibinfo {author} {\bibfnamefont {B.~A.~D.}\ \bibnamefont {Sukra}}, \ and\
  \bibinfo {author} {\bibfnamefont {G.}~\bibnamefont {Gabrielse}},\ }\href
  {\doibase 10.1103/PhysRevLett.130.071801} {\bibfield  {journal} {\bibinfo
  {journal} {Phys. Rev. Lett.}\ }\textbf {\bibinfo {volume} {130}},\ \bibinfo
  {pages} {071801} (\bibinfo {year} {2023})},\ \Eprint
  {http://arxiv.org/abs/2209.13084} {arXiv:2209.13084 [physics.atom-ph]}
  \BibitemShut {NoStop}%
\bibitem [{\citenamefont {Anastasi}\ \emph {et~al.}(2015)\citenamefont
  {Anastasi} \emph {et~al.}}]{Anastasi:2015qla}%
  \BibitemOpen
  \bibfield  {author} {\bibinfo {author} {\bibfnamefont {A.}~\bibnamefont
  {Anastasi}} \emph {et~al.},\ }\href {\doibase 10.1016/j.physletb.2015.10.003}
  {\bibfield  {journal} {\bibinfo  {journal} {Phys. Lett. B}\ }\textbf
  {\bibinfo {volume} {750}},\ \bibinfo {pages} {633} (\bibinfo {year}
  {2015})},\ \Eprint {http://arxiv.org/abs/1509.00740} {arXiv:1509.00740
  [hep-ex]} \BibitemShut {NoStop}%
\bibitem [{\citenamefont {Banerjee}\ \emph {et~al.}(2020)\citenamefont
  {Banerjee} \emph {et~al.}}]{NA64:2019auh}%
  \BibitemOpen
  \bibfield  {author} {\bibinfo {author} {\bibfnamefont {D.}~\bibnamefont
  {Banerjee}} \emph {et~al.} (\bibinfo {collaboration} {NA64}),\ }\href
  {\doibase 10.1103/PhysRevD.101.071101} {\bibfield  {journal} {\bibinfo
  {journal} {Phys. Rev. D}\ }\textbf {\bibinfo {volume} {101}},\ \bibinfo
  {pages} {071101} (\bibinfo {year} {2020})},\ \Eprint
  {http://arxiv.org/abs/1912.11389} {arXiv:1912.11389 [hep-ex]} \BibitemShut
  {NoStop}%
\bibitem [{\citenamefont {Depero}\ \emph {et~al.}(2020)\citenamefont {Depero}
  \emph {et~al.}}]{NA64:2020xxh}%
  \BibitemOpen
  \bibfield  {author} {\bibinfo {author} {\bibfnamefont {E.}~\bibnamefont
  {Depero}} \emph {et~al.} (\bibinfo {collaboration} {NA64}),\ }\href {\doibase
  10.1140/epjc/s10052-020-08725-x} {\bibfield  {journal} {\bibinfo  {journal}
  {Eur. Phys. J. C}\ }\textbf {\bibinfo {volume} {80}},\ \bibinfo {pages}
  {1159} (\bibinfo {year} {2020})},\ \Eprint {http://arxiv.org/abs/2009.02756}
  {arXiv:2009.02756 [hep-ex]} \BibitemShut {NoStop}%
\bibitem [{\citenamefont {Lees}\ \emph {et~al.}(2014)\citenamefont {Lees} \emph
  {et~al.}}]{BaBar:2014zli}%
  \BibitemOpen
  \bibfield  {author} {\bibinfo {author} {\bibfnamefont {J.~P.}\ \bibnamefont
  {Lees}} \emph {et~al.} (\bibinfo {collaboration} {BaBar}),\ }\href {\doibase
  10.1103/PhysRevLett.113.201801} {\bibfield  {journal} {\bibinfo  {journal}
  {Phys. Rev. Lett.}\ }\textbf {\bibinfo {volume} {113}},\ \bibinfo {pages}
  {201801} (\bibinfo {year} {2014})},\ \Eprint {http://arxiv.org/abs/1406.2980}
  {arXiv:1406.2980 [hep-ex]} \BibitemShut {NoStop}%
\bibitem [{\citenamefont {Krasznahorkay}\ \emph {et~al.}(2019)\citenamefont
  {Krasznahorkay} \emph {et~al.}}]{Krasznahorkay:2019lyl}%
  \BibitemOpen
  \bibfield  {author} {\bibinfo {author} {\bibfnamefont {A.~J.}\ \bibnamefont
  {Krasznahorkay}} \emph {et~al.},\ }\href@noop {} {\  (\bibinfo {year}
  {2019})},\ \Eprint {http://arxiv.org/abs/1910.10459} {arXiv:1910.10459
  [nucl-ex]} \BibitemShut {NoStop}%
\bibitem [{\citenamefont {Dine}\ and\ \citenamefont
  {Seiberg}(1986)}]{Dine:1986bg}%
  \BibitemOpen
  \bibfield  {author} {\bibinfo {author} {\bibfnamefont {M.}~\bibnamefont
  {Dine}}\ and\ \bibinfo {author} {\bibfnamefont {N.}~\bibnamefont {Seiberg}},\
  }\href {\doibase 10.1016/0550-3213(86)90043-X} {\bibfield  {journal}
  {\bibinfo  {journal} {Nucl. Phys. B}\ }\textbf {\bibinfo {volume} {273}},\
  \bibinfo {pages} {109} (\bibinfo {year} {1986})}\BibitemShut {NoStop}%
\bibitem [{\citenamefont {Barr}\ and\ \citenamefont
  {Seckel}(1992)}]{Barr:1992qq}%
  \BibitemOpen
  \bibfield  {author} {\bibinfo {author} {\bibfnamefont {S.~M.}\ \bibnamefont
  {Barr}}\ and\ \bibinfo {author} {\bibfnamefont {D.}~\bibnamefont {Seckel}},\
  }\href {\doibase 10.1103/PhysRevD.46.539} {\bibfield  {journal} {\bibinfo
  {journal} {Phys. Rev. D}\ }\textbf {\bibinfo {volume} {46}},\ \bibinfo
  {pages} {539} (\bibinfo {year} {1992})}\BibitemShut {NoStop}%
\bibitem [{\citenamefont {Kamionkowski}\ and\ \citenamefont
  {March-Russell}(1992{\natexlab{a}})}]{Kamionkowski:1992mf}%
  \BibitemOpen
  \bibfield  {author} {\bibinfo {author} {\bibfnamefont {M.}~\bibnamefont
  {Kamionkowski}}\ and\ \bibinfo {author} {\bibfnamefont {J.}~\bibnamefont
  {March-Russell}},\ }\href {\doibase 10.1016/0370-2693(92)90492-M} {\bibfield
  {journal} {\bibinfo  {journal} {Phys. Lett. B}\ }\textbf {\bibinfo {volume}
  {282}},\ \bibinfo {pages} {137} (\bibinfo {year} {1992}{\natexlab{a}})},\
  \Eprint {http://arxiv.org/abs/hep-th/9202003} {arXiv:hep-th/9202003}
  \BibitemShut {NoStop}%
\bibitem [{\citenamefont {Kamionkowski}\ and\ \citenamefont
  {March-Russell}(1992{\natexlab{b}})}]{Kamionkowski:1992ax}%
  \BibitemOpen
  \bibfield  {author} {\bibinfo {author} {\bibfnamefont {M.}~\bibnamefont
  {Kamionkowski}}\ and\ \bibinfo {author} {\bibfnamefont {J.}~\bibnamefont
  {March-Russell}},\ }\href {\doibase 10.1103/PhysRevLett.69.1485} {\bibfield
  {journal} {\bibinfo  {journal} {Phys. Rev. Lett.}\ }\textbf {\bibinfo
  {volume} {69}},\ \bibinfo {pages} {1485} (\bibinfo {year}
  {1992}{\natexlab{b}})},\ \Eprint {http://arxiv.org/abs/hep-th/9201063}
  {arXiv:hep-th/9201063} \BibitemShut {NoStop}%
\bibitem [{\citenamefont {Holman}\ \emph {et~al.}(1992)\citenamefont {Holman},
  \citenamefont {Hsu}, \citenamefont {Kephart}, \citenamefont {Kolb},
  \citenamefont {Watkins},\ and\ \citenamefont {Widrow}}]{Holman:1992us}%
  \BibitemOpen
  \bibfield  {author} {\bibinfo {author} {\bibfnamefont {R.}~\bibnamefont
  {Holman}}, \bibinfo {author} {\bibfnamefont {S.~D.~H.}\ \bibnamefont {Hsu}},
  \bibinfo {author} {\bibfnamefont {T.~W.}\ \bibnamefont {Kephart}}, \bibinfo
  {author} {\bibfnamefont {E.~W.}\ \bibnamefont {Kolb}}, \bibinfo {author}
  {\bibfnamefont {R.}~\bibnamefont {Watkins}}, \ and\ \bibinfo {author}
  {\bibfnamefont {L.~M.}\ \bibnamefont {Widrow}},\ }\href {\doibase
  10.1016/0370-2693(92)90491-L} {\bibfield  {journal} {\bibinfo  {journal}
  {Phys. Lett. B}\ }\textbf {\bibinfo {volume} {282}},\ \bibinfo {pages} {132}
  (\bibinfo {year} {1992})},\ \Eprint {http://arxiv.org/abs/hep-ph/9203206}
  {arXiv:hep-ph/9203206} \BibitemShut {NoStop}%
\bibitem [{\citenamefont {Kallosh}\ \emph {et~al.}(1995)\citenamefont
  {Kallosh}, \citenamefont {Linde}, \citenamefont {Linde},\ and\ \citenamefont
  {Susskind}}]{Kallosh:1995hi}%
  \BibitemOpen
  \bibfield  {author} {\bibinfo {author} {\bibfnamefont {R.}~\bibnamefont
  {Kallosh}}, \bibinfo {author} {\bibfnamefont {A.~D.}\ \bibnamefont {Linde}},
  \bibinfo {author} {\bibfnamefont {D.~A.}\ \bibnamefont {Linde}}, \ and\
  \bibinfo {author} {\bibfnamefont {L.}~\bibnamefont {Susskind}},\ }\href
  {\doibase 10.1103/PhysRevD.52.912} {\bibfield  {journal} {\bibinfo  {journal}
  {Phys. Rev. D}\ }\textbf {\bibinfo {volume} {52}},\ \bibinfo {pages} {912}
  (\bibinfo {year} {1995})},\ \Eprint {http://arxiv.org/abs/hep-th/9502069}
  {arXiv:hep-th/9502069} \BibitemShut {NoStop}%
\bibitem [{\citenamefont {Carpenter}\ \emph
  {et~al.}(2009{\natexlab{a}})\citenamefont {Carpenter}, \citenamefont {Dine},\
  and\ \citenamefont {Festuccia}}]{Carpenter:2009zs}%
  \BibitemOpen
  \bibfield  {author} {\bibinfo {author} {\bibfnamefont {L.~M.}\ \bibnamefont
  {Carpenter}}, \bibinfo {author} {\bibfnamefont {M.}~\bibnamefont {Dine}}, \
  and\ \bibinfo {author} {\bibfnamefont {G.}~\bibnamefont {Festuccia}},\ }\href
  {\doibase 10.1103/PhysRevD.80.125017} {\bibfield  {journal} {\bibinfo
  {journal} {Phys. Rev. D}\ }\textbf {\bibinfo {volume} {80}},\ \bibinfo
  {pages} {125017} (\bibinfo {year} {2009}{\natexlab{a}})},\ \Eprint
  {http://arxiv.org/abs/0906.1273} {arXiv:0906.1273 [hep-th]} \BibitemShut
  {NoStop}%
\bibitem [{\citenamefont {Carpenter}\ \emph
  {et~al.}(2009{\natexlab{b}})\citenamefont {Carpenter}, \citenamefont {Dine},
  \citenamefont {Festuccia},\ and\ \citenamefont {Ubaldi}}]{Carpenter:2009sw}%
  \BibitemOpen
  \bibfield  {author} {\bibinfo {author} {\bibfnamefont {L.~M.}\ \bibnamefont
  {Carpenter}}, \bibinfo {author} {\bibfnamefont {M.}~\bibnamefont {Dine}},
  \bibinfo {author} {\bibfnamefont {G.}~\bibnamefont {Festuccia}}, \ and\
  \bibinfo {author} {\bibfnamefont {L.}~\bibnamefont {Ubaldi}},\ }\href
  {\doibase 10.1103/PhysRevD.80.125023} {\bibfield  {journal} {\bibinfo
  {journal} {Phys. Rev. D}\ }\textbf {\bibinfo {volume} {80}},\ \bibinfo
  {pages} {125023} (\bibinfo {year} {2009}{\natexlab{b}})},\ \Eprint
  {http://arxiv.org/abs/0906.5015} {arXiv:0906.5015 [hep-th]} \BibitemShut
  {NoStop}%
\bibitem [{\citenamefont {Girmohanta}\ \emph {et~al.}(2023)\citenamefont
  {Girmohanta}, \citenamefont {Qiu}, \citenamefont {Wang},\ and\ \citenamefont
  {Yanagida}}]{Girmohanta:2023ghm}%
  \BibitemOpen
  \bibfield  {author} {\bibinfo {author} {\bibfnamefont {S.}~\bibnamefont
  {Girmohanta}}, \bibinfo {author} {\bibfnamefont {Y.-C.}\ \bibnamefont {Qiu}},
  \bibinfo {author} {\bibfnamefont {J.-W.}\ \bibnamefont {Wang}}, \ and\
  \bibinfo {author} {\bibfnamefont {T.~T.}\ \bibnamefont {Yanagida}},\ }\href
  {\doibase 10.1103/PhysRevD.108.015028} {\bibfield  {journal} {\bibinfo
  {journal} {Phys. Rev. D}\ }\textbf {\bibinfo {volume} {108}},\ \bibinfo
  {pages} {015028} (\bibinfo {year} {2023})},\ \Eprint
  {http://arxiv.org/abs/2303.02852} {arXiv:2303.02852 [hep-ph]} \BibitemShut
  {NoStop}%
\bibitem [{\citenamefont {Baker}\ \emph {et~al.}(2006)\citenamefont {Baker}
  \emph {et~al.}}]{Baker:2006ts}%
  \BibitemOpen
  \bibfield  {author} {\bibinfo {author} {\bibfnamefont {C.~A.}\ \bibnamefont
  {Baker}} \emph {et~al.},\ }\href {\doibase 10.1103/PhysRevLett.97.131801}
  {\bibfield  {journal} {\bibinfo  {journal} {Phys. Rev. Lett.}\ }\textbf
  {\bibinfo {volume} {97}},\ \bibinfo {pages} {131801} (\bibinfo {year}
  {2006})},\ \Eprint {http://arxiv.org/abs/hep-ex/0602020}
  {arXiv:hep-ex/0602020} \BibitemShut {NoStop}%
\bibitem [{\citenamefont {Pendlebury}\ \emph {et~al.}(2015)\citenamefont
  {Pendlebury} \emph {et~al.}}]{Pendlebury:2015lrz}%
  \BibitemOpen
  \bibfield  {author} {\bibinfo {author} {\bibfnamefont {J.~M.}\ \bibnamefont
  {Pendlebury}} \emph {et~al.},\ }\href {\doibase 10.1103/PhysRevD.92.092003}
  {\bibfield  {journal} {\bibinfo  {journal} {Phys. Rev. D}\ }\textbf {\bibinfo
  {volume} {92}},\ \bibinfo {pages} {092003} (\bibinfo {year} {2015})},\
  \Eprint {http://arxiv.org/abs/1509.04411} {arXiv:1509.04411 [hep-ex]}
  \BibitemShut {NoStop}%
\bibitem [{\citenamefont {Riordan}\ \emph {et~al.}(1987)\citenamefont {Riordan}
  \emph {et~al.}}]{Riordan:1987aw}%
  \BibitemOpen
  \bibfield  {author} {\bibinfo {author} {\bibfnamefont {E.~M.}\ \bibnamefont
  {Riordan}} \emph {et~al.},\ }\href {\doibase 10.1103/PhysRevLett.59.755}
  {\bibfield  {journal} {\bibinfo  {journal} {Phys. Rev. Lett.}\ }\textbf
  {\bibinfo {volume} {59}},\ \bibinfo {pages} {755} (\bibinfo {year}
  {1987})}\BibitemShut {NoStop}%
\bibitem [{\citenamefont {Kibble}(1976)}]{Kibble:1976sj}%
  \BibitemOpen
  \bibfield  {author} {\bibinfo {author} {\bibfnamefont {T.~W.~B.}\
  \bibnamefont {Kibble}},\ }\href {\doibase 10.1088/0305-4470/9/8/029}
  {\bibfield  {journal} {\bibinfo  {journal} {J. Phys. A}\ }\textbf {\bibinfo
  {volume} {9}},\ \bibinfo {pages} {1387} (\bibinfo {year} {1976})}\BibitemShut
  {NoStop}%
\bibitem [{\citenamefont {Kibble}\ \emph {et~al.}(1982)\citenamefont {Kibble},
  \citenamefont {Lazarides},\ and\ \citenamefont {Shafi}}]{Kibble:1982dd}%
  \BibitemOpen
  \bibfield  {author} {\bibinfo {author} {\bibfnamefont {T.~W.~B.}\
  \bibnamefont {Kibble}}, \bibinfo {author} {\bibfnamefont {G.}~\bibnamefont
  {Lazarides}}, \ and\ \bibinfo {author} {\bibfnamefont {Q.}~\bibnamefont
  {Shafi}},\ }\href {\doibase 10.1103/PhysRevD.26.435} {\bibfield  {journal}
  {\bibinfo  {journal} {Phys. Rev. D}\ }\textbf {\bibinfo {volume} {26}},\
  \bibinfo {pages} {435} (\bibinfo {year} {1982})}\BibitemShut {NoStop}%
\bibitem [{\citenamefont {Beyer}\ and\ \citenamefont
  {Sarkar}(2023)}]{Beyer:2022ywc}%
  \BibitemOpen
  \bibfield  {author} {\bibinfo {author} {\bibfnamefont {K.~A.}\ \bibnamefont
  {Beyer}}\ and\ \bibinfo {author} {\bibfnamefont {S.}~\bibnamefont {Sarkar}},\
  }\href {\doibase 10.21468/SciPostPhys.15.1.003} {\bibfield  {journal}
  {\bibinfo  {journal} {SciPost Phys.}\ }\textbf {\bibinfo {volume} {15}},\
  \bibinfo {pages} {003} (\bibinfo {year} {2023})},\ \Eprint
  {http://arxiv.org/abs/2211.14635} {arXiv:2211.14635 [hep-ph]} \BibitemShut
  {NoStop}%
\bibitem [{\citenamefont {Dine}(2023)}]{Dine:2023qsq}%
  \BibitemOpen
  \bibfield  {author} {\bibinfo {author} {\bibfnamefont {M.}~\bibnamefont
  {Dine}},\ }\href@noop {} {\  (\bibinfo {year} {2023})},\ \Eprint
  {http://arxiv.org/abs/2307.04710} {arXiv:2307.04710 [hep-ph]} \BibitemShut
  {NoStop}%
\bibitem [{\citenamefont {Vilenkin}(1981)}]{Vilenkin:1981zs}%
  \BibitemOpen
  \bibfield  {author} {\bibinfo {author} {\bibfnamefont {A.}~\bibnamefont
  {Vilenkin}},\ }\href {\doibase 10.1103/PhysRevD.23.852} {\bibfield  {journal}
  {\bibinfo  {journal} {Phys. Rev. D}\ }\textbf {\bibinfo {volume} {23}},\
  \bibinfo {pages} {852} (\bibinfo {year} {1981})}\BibitemShut {NoStop}%
\bibitem [{\citenamefont {Gelmini}\ \emph {et~al.}(1989)\citenamefont
  {Gelmini}, \citenamefont {Gleiser},\ and\ \citenamefont
  {Kolb}}]{Gelmini:1988sf}%
  \BibitemOpen
  \bibfield  {author} {\bibinfo {author} {\bibfnamefont {G.~B.}\ \bibnamefont
  {Gelmini}}, \bibinfo {author} {\bibfnamefont {M.}~\bibnamefont {Gleiser}}, \
  and\ \bibinfo {author} {\bibfnamefont {E.~W.}\ \bibnamefont {Kolb}},\ }\href
  {\doibase 10.1103/PhysRevD.39.1558} {\bibfield  {journal} {\bibinfo
  {journal} {Phys. Rev. D}\ }\textbf {\bibinfo {volume} {39}},\ \bibinfo
  {pages} {1558} (\bibinfo {year} {1989})}\BibitemShut {NoStop}%
\bibitem [{\citenamefont {Larsson}\ \emph {et~al.}(1997)\citenamefont
  {Larsson}, \citenamefont {Sarkar},\ and\ \citenamefont
  {White}}]{Larsson:1996sp}%
  \BibitemOpen
  \bibfield  {author} {\bibinfo {author} {\bibfnamefont {S.~E.}\ \bibnamefont
  {Larsson}}, \bibinfo {author} {\bibfnamefont {S.}~\bibnamefont {Sarkar}}, \
  and\ \bibinfo {author} {\bibfnamefont {P.~L.}\ \bibnamefont {White}},\ }\href
  {\doibase 10.1103/PhysRevD.55.5129} {\bibfield  {journal} {\bibinfo
  {journal} {Phys. Rev. D}\ }\textbf {\bibinfo {volume} {55}},\ \bibinfo
  {pages} {5129} (\bibinfo {year} {1997})},\ \Eprint
  {http://arxiv.org/abs/hep-ph/9608319} {arXiv:hep-ph/9608319} \BibitemShut
  {NoStop}%
\bibitem [{\citenamefont {Kawasaki}\ \emph {et~al.}(2015)\citenamefont
  {Kawasaki}, \citenamefont {Saikawa},\ and\ \citenamefont
  {Sekiguchi}}]{Kawasaki:2014sqa}%
  \BibitemOpen
  \bibfield  {author} {\bibinfo {author} {\bibfnamefont {M.}~\bibnamefont
  {Kawasaki}}, \bibinfo {author} {\bibfnamefont {K.}~\bibnamefont {Saikawa}}, \
  and\ \bibinfo {author} {\bibfnamefont {T.}~\bibnamefont {Sekiguchi}},\ }\href
  {\doibase 10.1103/PhysRevD.91.065014} {\bibfield  {journal} {\bibinfo
  {journal} {Phys. Rev. D}\ }\textbf {\bibinfo {volume} {91}},\ \bibinfo
  {pages} {065014} (\bibinfo {year} {2015})},\ \Eprint
  {http://arxiv.org/abs/1412.0789} {arXiv:1412.0789 [hep-ph]} \BibitemShut
  {NoStop}%
\bibitem [{\citenamefont {Huang}\ and\ \citenamefont
  {Sikivie}(1985)}]{Huang:1985tt}%
  \BibitemOpen
  \bibfield  {author} {\bibinfo {author} {\bibfnamefont {M.~C.}\ \bibnamefont
  {Huang}}\ and\ \bibinfo {author} {\bibfnamefont {P.}~\bibnamefont
  {Sikivie}},\ }\href {\doibase 10.1103/PhysRevD.32.1560} {\bibfield  {journal}
  {\bibinfo  {journal} {Phys. Rev. D}\ }\textbf {\bibinfo {volume} {32}},\
  \bibinfo {pages} {1560} (\bibinfo {year} {1985})}\BibitemShut {NoStop}%
\bibitem [{\citenamefont {Hiramatsu}\ \emph {et~al.}(2013)\citenamefont
  {Hiramatsu}, \citenamefont {Kawasaki}, \citenamefont {Saikawa},\ and\
  \citenamefont {Sekiguchi}}]{Hiramatsu:2012sc}%
  \BibitemOpen
  \bibfield  {author} {\bibinfo {author} {\bibfnamefont {T.}~\bibnamefont
  {Hiramatsu}}, \bibinfo {author} {\bibfnamefont {M.}~\bibnamefont {Kawasaki}},
  \bibinfo {author} {\bibfnamefont {K.}~\bibnamefont {Saikawa}}, \ and\
  \bibinfo {author} {\bibfnamefont {T.}~\bibnamefont {Sekiguchi}},\ }\href
  {\doibase 10.1088/1475-7516/2013/01/001} {\bibfield  {journal} {\bibinfo
  {journal} {JCAP}\ }\textbf {\bibinfo {volume} {01}},\ \bibinfo {pages} {001}
  (\bibinfo {year} {2013})},\ \Eprint {http://arxiv.org/abs/1207.3166}
  {arXiv:1207.3166 [hep-ph]} \BibitemShut {NoStop}%
\bibitem [{\citenamefont {Roussy}\ \emph {et~al.}(2023)\citenamefont {Roussy},
  \citenamefont {Caldwell}, \citenamefont {Wright}, \citenamefont {Cairncross},
  \citenamefont {Shagam}, \citenamefont {Ng}, \citenamefont {Schlossberger},
  \citenamefont {Park}, \citenamefont {Wang}, \citenamefont {Ye},\ and\
  \citenamefont {Cornell}}]{Roussy_2023}%
  \BibitemOpen
  \bibfield  {author} {\bibinfo {author} {\bibfnamefont {T.~S.}\ \bibnamefont
  {Roussy}}, \bibinfo {author} {\bibfnamefont {L.}~\bibnamefont {Caldwell}},
  \bibinfo {author} {\bibfnamefont {T.}~\bibnamefont {Wright}}, \bibinfo
  {author} {\bibfnamefont {W.~B.}\ \bibnamefont {Cairncross}}, \bibinfo
  {author} {\bibfnamefont {Y.}~\bibnamefont {Shagam}}, \bibinfo {author}
  {\bibfnamefont {K.~B.}\ \bibnamefont {Ng}}, \bibinfo {author} {\bibfnamefont
  {N.}~\bibnamefont {Schlossberger}}, \bibinfo {author} {\bibfnamefont {S.~Y.}\
  \bibnamefont {Park}}, \bibinfo {author} {\bibfnamefont {A.}~\bibnamefont
  {Wang}}, \bibinfo {author} {\bibfnamefont {J.}~\bibnamefont {Ye}}, \ and\
  \bibinfo {author} {\bibfnamefont {E.~A.}\ \bibnamefont {Cornell}},\ }\href
  {\doibase 10.1126/science.adg4084} {\bibfield  {journal} {\bibinfo  {journal}
  {Science}\ }\textbf {\bibinfo {volume} {381}},\ \bibinfo {pages} {46–50}
  (\bibinfo {year} {2023})}\BibitemShut {NoStop}%
\bibitem [{\citenamefont {Barr}\ and\ \citenamefont {Zee}(1990)}]{Barr:1990vd}%
  \BibitemOpen
  \bibfield  {author} {\bibinfo {author} {\bibfnamefont {S.~M.}\ \bibnamefont
  {Barr}}\ and\ \bibinfo {author} {\bibfnamefont {A.}~\bibnamefont {Zee}},\
  }\href {\doibase 10.1103/PhysRevLett.65.21} {\bibfield  {journal} {\bibinfo
  {journal} {Phys. Rev. Lett.}\ }\textbf {\bibinfo {volume} {65}},\ \bibinfo
  {pages} {21} (\bibinfo {year} {1990})},\ \bibinfo {note} {[Erratum:
  Phys.Rev.Lett. 65, 2920 (1990)]}\BibitemShut {NoStop}%
\bibitem [{\citenamefont {Nakai}\ and\ \citenamefont
  {Reece}(2017)}]{Nakai:2016atk}%
  \BibitemOpen
  \bibfield  {author} {\bibinfo {author} {\bibfnamefont {Y.}~\bibnamefont
  {Nakai}}\ and\ \bibinfo {author} {\bibfnamefont {M.}~\bibnamefont {Reece}},\
  }\href {\doibase 10.1007/JHEP08(2017)031} {\bibfield  {journal} {\bibinfo
  {journal} {JHEP}\ }\textbf {\bibinfo {volume} {08}},\ \bibinfo {pages} {031}
  (\bibinfo {year} {2017})},\ \Eprint {http://arxiv.org/abs/1612.08090}
  {arXiv:1612.08090 [hep-ph]} \BibitemShut {NoStop}%
\bibitem [{\citenamefont {Cesarotti}\ \emph {et~al.}(2019)\citenamefont
  {Cesarotti}, \citenamefont {Lu}, \citenamefont {Nakai}, \citenamefont
  {Parikh},\ and\ \citenamefont {Reece}}]{Cesarotti:2018huy}%
  \BibitemOpen
  \bibfield  {author} {\bibinfo {author} {\bibfnamefont {C.}~\bibnamefont
  {Cesarotti}}, \bibinfo {author} {\bibfnamefont {Q.}~\bibnamefont {Lu}},
  \bibinfo {author} {\bibfnamefont {Y.}~\bibnamefont {Nakai}}, \bibinfo
  {author} {\bibfnamefont {A.}~\bibnamefont {Parikh}}, \ and\ \bibinfo {author}
  {\bibfnamefont {M.}~\bibnamefont {Reece}},\ }\href {\doibase
  10.1007/JHEP05(2019)059} {\bibfield  {journal} {\bibinfo  {journal} {JHEP}\
  }\textbf {\bibinfo {volume} {05}},\ \bibinfo {pages} {059} (\bibinfo {year}
  {2019})},\ \Eprint {http://arxiv.org/abs/1810.07736} {arXiv:1810.07736
  [hep-ph]} \BibitemShut {NoStop}%
\bibitem [{\citenamefont {Graner}\ \emph {et~al.}(2016)\citenamefont {Graner},
  \citenamefont {Chen}, \citenamefont {Lindahl},\ and\ \citenamefont
  {Heckel}}]{Graner_2016}%
  \BibitemOpen
  \bibfield  {author} {\bibinfo {author} {\bibfnamefont {B.}~\bibnamefont
  {Graner}}, \bibinfo {author} {\bibfnamefont {Y.}~\bibnamefont {Chen}},
  \bibinfo {author} {\bibfnamefont {E.}~\bibnamefont {Lindahl}}, \ and\
  \bibinfo {author} {\bibfnamefont {B.}~\bibnamefont {Heckel}},\ }\href
  {\doibase 10.1103/physrevlett.116.161601} {\bibfield  {journal} {\bibinfo
  {journal} {Physical Review Letters}\ }\textbf {\bibinfo {volume} {116}}
  (\bibinfo {year} {2016}),\ 10.1103/physrevlett.116.161601}\BibitemShut
  {NoStop}%
\bibitem [{\citenamefont {Batell}\ \emph {et~al.}(2011)\citenamefont {Batell},
  \citenamefont {Pospelov},\ and\ \citenamefont {Ritz}}]{Batell:2009jf}%
  \BibitemOpen
  \bibfield  {author} {\bibinfo {author} {\bibfnamefont {B.}~\bibnamefont
  {Batell}}, \bibinfo {author} {\bibfnamefont {M.}~\bibnamefont {Pospelov}}, \
  and\ \bibinfo {author} {\bibfnamefont {A.}~\bibnamefont {Ritz}},\ }\href
  {\doibase 10.1103/PhysRevD.83.054005} {\bibfield  {journal} {\bibinfo
  {journal} {Phys. Rev. D}\ }\textbf {\bibinfo {volume} {83}},\ \bibinfo
  {pages} {054005} (\bibinfo {year} {2011})},\ \Eprint
  {http://arxiv.org/abs/0911.4938} {arXiv:0911.4938 [hep-ph]} \BibitemShut
  {NoStop}%
\bibitem [{\citenamefont {Aaij}\ \emph {et~al.}(2015)\citenamefont {Aaij} \emph
  {et~al.}}]{LHCb:2015ycz}%
  \BibitemOpen
  \bibfield  {author} {\bibinfo {author} {\bibfnamefont {R.}~\bibnamefont
  {Aaij}} \emph {et~al.} (\bibinfo {collaboration} {LHCb}),\ }\href {\doibase
  10.1007/JHEP04(2015)064} {\bibfield  {journal} {\bibinfo  {journal} {JHEP}\
  }\textbf {\bibinfo {volume} {04}},\ \bibinfo {pages} {064} (\bibinfo {year}
  {2015})},\ \Eprint {http://arxiv.org/abs/1501.03038} {arXiv:1501.03038
  [hep-ex]} \BibitemShut {NoStop}%
\bibitem [{\citenamefont {J\"ager}\ and\ \citenamefont
  {Martin~Camalich}(2013)}]{Jager:2012uw}%
  \BibitemOpen
  \bibfield  {author} {\bibinfo {author} {\bibfnamefont {S.}~\bibnamefont
  {J\"ager}}\ and\ \bibinfo {author} {\bibfnamefont {J.}~\bibnamefont
  {Martin~Camalich}},\ }\href {\doibase 10.1007/JHEP05(2013)043} {\bibfield
  {journal} {\bibinfo  {journal} {JHEP}\ }\textbf {\bibinfo {volume} {05}},\
  \bibinfo {pages} {043} (\bibinfo {year} {2013})},\ \Eprint
  {http://arxiv.org/abs/1212.2263} {arXiv:1212.2263 [hep-ph]} \BibitemShut
  {NoStop}%
\bibitem [{\citenamefont {Wei}\ \emph {et~al.}(2009)\citenamefont {Wei} \emph
  {et~al.}}]{Belle:2009zue}%
  \BibitemOpen
  \bibfield  {author} {\bibinfo {author} {\bibfnamefont {J.~T.}\ \bibnamefont
  {Wei}} \emph {et~al.} (\bibinfo {collaboration} {Belle}),\ }\href {\doibase
  10.1103/PhysRevLett.103.171801} {\bibfield  {journal} {\bibinfo  {journal}
  {Phys. Rev. Lett.}\ }\textbf {\bibinfo {volume} {103}},\ \bibinfo {pages}
  {171801} (\bibinfo {year} {2009})},\ \Eprint {http://arxiv.org/abs/0904.0770}
  {arXiv:0904.0770 [hep-ex]} \BibitemShut {NoStop}%
\bibitem [{\citenamefont {Aubert}\ \emph {et~al.}(2009)\citenamefont {Aubert}
  \emph {et~al.}}]{BaBar:2008jdv}%
  \BibitemOpen
  \bibfield  {author} {\bibinfo {author} {\bibfnamefont {B.}~\bibnamefont
  {Aubert}} \emph {et~al.} (\bibinfo {collaboration} {BaBar}),\ }\href
  {\doibase 10.1103/PhysRevLett.102.091803} {\bibfield  {journal} {\bibinfo
  {journal} {Phys. Rev. Lett.}\ }\textbf {\bibinfo {volume} {102}},\ \bibinfo
  {pages} {091803} (\bibinfo {year} {2009})},\ \Eprint
  {http://arxiv.org/abs/0807.4119} {arXiv:0807.4119 [hep-ex]} \BibitemShut
  {NoStop}%
\end{thebibliography}%

\end{document}